\documentclass[aps,prd,preprint,superscriptaddress]{revtex4-1}
\usepackage{amsmath,amssymb}
\usepackage{graphicx}
\makeatletter

\@addtoreset{equation}{section}
\makeatother

\begin{document}
\preprint{OU-HET-967}

\title{
  Lattice computation of the Dirac eigenvalue density 
  in the perturbative regime of QCD
}


\author{Katsumasa Nakayama}
\email[]{katumasa@post.kek.jp}
\affiliation{Department of Physics, 
  Nagoya University, Nagoya, 464-8602, Japan}
\affiliation{KEK Theory Center, 
  High Energy Accelerator Research Organization (KEK), 
  Tsukuba 305-0801, Japan}

\author{Hidenori Fukaya}
\affiliation{Department of Physics, 
  Osaka University, Toyonaka 560-0043, Japan}

\author{Shoji Hashimoto}
\affiliation{KEK Theory Center, 
  High Energy Accelerator Research Organization (KEK), 
  Tsukuba 305-0801, Japan}
\affiliation{School of High Energy Accelerator Science, 
  The Graduate University for Advanced Studies (Sokendai),
  Tsukuba 305-0801, Japan
}


%
%
\date{\today}
\begin{abstract}
  The eigenvalue spectrum $\rho(\lambda)$ of the Dirac operator is
  numerically calculated in lattice QCD with 2+1 flavors of dynamical
  domain-wall fermions. 
  In the high-energy regime, the discretization effects become
 significant. We subtract them at the leading order and then take
  the continuum limit with lattice data at three lattice spacings. 
  Lattice results for the exponent $\partial\ln\rho/\partial\ln\lambda$
  are matched to continuum perturbation theory, which is known up to
  $O(\alpha_s^4)$, to extract the strong coupling constant $\alpha_s$. 
\end{abstract}


\maketitle
\section{Introduction}
\label{sec:introduction}
The Dirac operator $D$ is a fundamental building block of gauge theories 
such as Quantum Chromodynamics (QCD), the underlying theory of strong interaction.
It defines the interaction between quarks and the background gauge field
in a gauge invariant manner.
Any observable consisting of quarks in QCD can be written in terms of its eigenmodes,
{\it i.e.} eigenvalues and their associated eigenfunctions,
after an average over background gauge configurations with a weight
determined by the path-integral formulation.
The eigenvalues are gauge invariant and any quantity built from them
is invariant under gauge transformations.

The eigenvalue distribution of the Dirac operator reflects the
dynamics of QCD.
The best-known example is the Banks-Casher relation
\cite{Banks:1979yr}, which relates the near-zero eigenvalue density to the
chiral condensate $\langle\overline{\psi}\psi\rangle$, 
the order parameter of spontaneous chiral symmetry breaking in QCD.
In the lattice simulation of QCD, the determination of the chiral
condensate can therefore be performed by numerically counting the
number of near-zero eigenmodes on finite-volume lattices.
In order to correctly identify the near-zero eigenmodes, chiral
symmetry needs to be precisely realized in the lattice fermion
formulation. A previous work \cite{Fukaya:2009fh,Fukaya:2010na} 
utilized a chirally symmetric lattice operator, the so-called
overlap-Dirac operator, and explicitly calculated the individual
eigenvalues to obtain the near-zero eigenvalue density.
The eigenvalue density can also be estimated stochastically.
The first such attempt was performed using Wilson fermions
\cite{Giusti:2008vb}. More recently we applied a slightly different
stochastic approach to calculate the eigenvalue spectrum of
domain-wall fermions, which are also chiral lattice fermions,
and achieved a precise determination of the chiral
condensate \cite{Cossu:2016eqs}.

Apart from the near-zero eigenvalues, the eigenvalue spectrum
$\rho(\lambda)$ carries more information about the dynamics of the
system. 
In particular, its scale dependence $d\rho(\lambda)/d\lambda$ is
precisely estimated by perturbation theory at 
sufficiently high energy scales where the strong coupling constant $\alpha_s$ is small, {\it i.e.}
$\lambda\gg\Lambda_{\rm QCD}$, with $\Lambda_{\rm QCD}$ the QCD scale.
It is therefore of interest to test the perturbative expansion of QCD
against non-perturbatively calculated lattice results.
So far, tests of high-order perturbation theory against
non-perturbative lattice calculations have been performed for the vacuum
polarization function at the order of $\alpha_s^4$
\cite{Tomii:2016xiv,Tomii:2017cbt}
and for the charmonium correlation function at $\alpha_s^3$
\cite{Allison:2008xk,Nakayama:2016atf}.
Additionally, stochastic perturbation theory has been applied to
simple quantities such as the plaquette
\cite{Bali:2014fea} and the static quark self-energy
\cite{Bali:2013pla}.
In this work, we perform another such test at $\alpha_s^4$ for the
scale dependence of $\rho(\lambda)$.
Since the spectral function can be calculated very precisely on the 
lattice, it also serves as an input for the determination of
$\alpha_s$.

There have been a number of lattice studies of the spectral function
aiming at finding the so-called {\it conformal} regime of many-flavor
QCD or related models, see {\it e.g.}
\cite{Patella:2012da,Cheng:2013eu} for instance,
and \cite{Cichy:2013eoa} applied the same technique to two-flavor
QCD. 
For the systems that have conformal (or scale) invariance, the
scale dependence of the spectral density may be parametrized by a
constant $\gamma_*$, which corresponds to the mass anomalous dimension
of associated fermions.
For QCD, for which the coupling constant runs and eventually diverges
in the low-energy region, this relation does not hold beyond the
one-loop level.
See the discussions in Section~\ref{sec:Ano_dim} for details.


This paper presents a lattice calculation of the Dirac spectral
density in the perturbative regime. 
We calculate the eigenvalue density in the whole energy range from
zero up to the lattice cutoff with the domain-wall fermion formulation
using a stochastic technique to evaluate the average number of
eigenvalues in small intervals.
The lattice results for the eigenvalue density are then extrapolated
to the continuum limit using data at three lattice spacings
in the range $a\simeq$ 0.044--0.080~fm.
We then investigate the consistency with continuum perturbation
theory, which is available up to $O(\alpha_s^4)$.
We also attempt to extract $\alpha_s$ using the spectral function as
an input. 

Since we are interested in a relatively high-energy region
where the Dirac eigenvalue $\lambda$ is not much smaller than the
lattice cutoff $1/a$,
discretization effects need to be eliminated
as far as possible, in order to achieve precise results.
At tree level we calculate the eigenvalue of the Dirac operator
constructed from domain-wall fermions and identify
discretization effects.
It turns out that discretization effects can be significantly
reduced by choosing a value of Pauli-Villars mass different from
its standard value.
Domain-wall fermions are first defined in five-dimensional space and
the eigenmodes on their four-dimensional surface are taken as the
physical fermion modes.
Our construction corresponds to a slightly different prescription to
cancel the bulk effects of the five-dimensional fermion modes
leaving the physical modes on the four-dimensional surface.
Using this scheme, the remaining discretization effects are made small,
so that we are able to extrapolate the lattice data to the continuum
limit with a linear ansatz in $a^2$.


The rest of this paper is organized as follows.
In Section~\ref{sec:Ano_dim} we briefly discuss the perturbative
results for the Dirac spectral density in QCD.
It contains a review of perturbative results for the spectral function
$\rho(\lambda)$ and its exponent.
A discussion of the discretization effects for domain-wall fermions
is given in Section~\ref{sec:Tree_level}.
We then describe the lattice setup as well as the method to calculate
the eigenvalue density in Section~\ref{sec:lattice_details}.
The lattice results are presented and compared with the perturbative
expansion in Section~\ref{sec:eigen_lat}.
The extraction of the strong coupling constant is discussed in Section~\ref{sec:alpha_s}.
Our conclusions are in Section \ref{sec:conclusion}.
Some details of the perturbative coefficients are in 
Appendix~\ref{app:perturbative}, and a discussion of the locality of
the modified Dirac operator we used in this work is found in 
Appendix~\ref{sec:locality_and_chirality}.

A preliminary version of this work was presented in
\cite{Nakayama:2017led}.

\section{Dirac eigenvalue density}
\label{sec:Ano_dim}

The eigenvalue density $\rho(\lambda)$ of the Dirac operator $D$ is
defined as
\begin{equation}
  \label{eq:def_rho}
  \rho(\lambda) = \frac{1}{V} \left\langle
    \sum_k \delta(\lambda-\lambda_k) 
    \right\rangle,
\end{equation}
where $\langle\cdots\rangle$ stands for an average over gauge field
configurations.
The eigenvalue $\lambda_k$ of $D$ depends on the background gauge field.
In the free quark limit, $\lambda_k$ may be labeled by the four-momentum
$p_\mu$ as $|\lambda_k|^2 = p_\mu^2$,
and the eigenvalue density is given by a surface of a three-dimensional
sphere in momentum space:
$\rho_{\mathrm{free}}(\lambda)=N_c(3/4\pi^2)|\lambda|^3$.
($N_c$ is the number of colors, $N_c=3$.)
In theories that exibit conformal invariance, there exists a relation  
$\rho_{\mathrm{FP}}\sim |\lambda|^{4/(1+\gamma_*)-1}$
with $\gamma_*$ the mass anomalous dimension of the theory \cite{DelDebbio:2010ze}.
It also applies to the case where the theory approaches a
renormalization group fixed point.
This relation has been utilized to extract the mass anomalous dimension
at the fixed point in many-flavor QCD-like theories
\cite{Patella:2012da,Cheng:2013eu}.

In QCD, the spectral density has been perturbatively calculated in the
$\overline{\mathrm{MS}}$ scheme to order $\alpha_s^3$
\cite{Chetyrkin:1994ex,Kneur:2015dda} using
\begin{equation}
  \label{eq:rhoMSbar}
  \rho^{\mathrm{\overline{MS}}}(\lambda)=\frac{3|\lambda|^3}{4\pi^2}
  \left[
    1-\rho^{(1)}\frac{\alpha_s(\mu)}{\pi}
    -\rho^{(2)}\left(\frac{\alpha_s(\mu)}{\pi}\right)^2
    -\rho^{(3)}\left(\frac{\alpha_s(\mu)}{\pi}\right)^3
    +O\left(\left(\frac{\alpha_s(\mu)}{\pi}\right)^4\right)
  \right],
\end{equation}
where the coefficients $\rho^{(i)}$ depend on $\lambda$ and are
calculated as
\begin{eqnarray}
  \rho^{(1)} & =  & 8 \left(L_\lambda-\frac{5}{12}\right),
  \label{eq:rho1}
  \\
  \rho^{(2)} & = & 
  \frac{1}{8} \left[
    \frac{1}{2}\left(52 N_f-\frac{4406}{9}+\frac{32}{3}\zeta_3\right)
    \right.
  \label{eq:rho2}
  \nonumber\\ & & \;\;
  \left.
    -\frac{32}{9}(5 N_f-141)L_\lambda
    +\frac{32}{9}(2 N_f-81)\left(\frac{3}{2}L_\lambda^2-\frac{\pi^2}{8}\right)
  \right],
  \\
  \rho^{(3)} & = & 
  \frac{1}{32} \left[
    c_{40}\left(2L_\lambda^3-\frac{\pi^2}{2}L_\lambda\right)
    +c_{41}\left(\frac{3}{2}L_\lambda^2-\frac{\pi^2}{8}\right)
    +c_{42} L_\lambda + \frac{1}{2} c_{43} 
  \right].
  \label{eq:rho3}
\end{eqnarray}
Here, $L_\lambda\equiv\ln(\lambda/\mu)$ and $\mu$ denotes the
renormalization scale. 
The numerical constants are $\zeta_3\simeq 1.20205$ and
$c_{40}\simeq 4533.33$,
$c_{41}\simeq -11292.4$,
$c_{42}\simeq 12648.1$,
$c_{43}\simeq -15993.5$
for $N_f=3$ dynamical fermion flavors \cite{Kneur:2015dda}.
Numerical results for $N_f=2$ are also available in
\cite{Kneur:2015dda}. 
The Dirac eigenvalue $\lambda$ is renormalized in the same way as
the quark mass.
It is therefore scale dependent, and the scale is set to $\mu$, 
{\it i.e.} $\lambda(\mu)$.
Here and in the following sections, we suppress this scale
dependence for brevity.
$\lambda$ denotes its absolute value $|\lambda|$.
($\lambda$ is pure imaginary in the Euclidean continuum theory.)

One can reconstruct the scale-dependent coefficient of
the next order, {\it i.e.} $O(\alpha_s^4)$, through the
renormalization group equation.
Namely, we start from 
\begin{equation}
  \label{eq:RG}
  0 = \left[
    \frac{\partial}{\partial\ln\mu}
    -\gamma_m(\alpha_s)
    \left(1+\lambda\frac{\partial}{\partial\lambda}\right)
    +\beta(\alpha_s)
    \frac{\partial}{\partial\alpha_s}
  \right]
  \rho(\lambda),
\end{equation}
which follows from the scale invariance of the mode number,
an integral of the spectral density
$\int_0^M d\lambda\,\rho(\lambda)$ with an upper limit $M$
\cite{Giusti:2008vb}.
This can also be understood from the scale invariance of the scalar density
operator $m\bar{q}q$.
Since the spectral function is given by the chiral condensate
$\langle\bar{q}q\rangle$ with the valence quark mass set to an
imaginary value $i|\lambda|$, the renormalization group equation is
identical to that for $\bar{q}q$.
Here the beta function $\beta(\alpha_s)$ and mass anomalous dimension $\gamma_m(\alpha_s)$ are defined as
\begin{eqnarray}
  \beta(\alpha_s) & \equiv & 
  \frac{\partial\alpha_s}{\partial\ln{\mu}},
  \label{eq:beta}
  \\
  \gamma_m(\alpha_s) & \equiv & 
  -\frac{\partial\ln m(\mu)}{\partial\ln\mu},
  \label{eq:anodim}
\end{eqnarray}
and known up to $O(\alpha_s^6)$ and $O(\alpha_s^5)$, respectively.
Their explicit forms are summarized in Appendix~\ref{app:perturbative}
for convenience.
Since the $\mu$-dependence of $\rho(\lambda)/\lambda^3$ appears only through the
form $L_\lambda=\ln\lambda/\mu$, we may rewrite (\ref{eq:RG}) as
\begin{equation}
  (1+\gamma_m)\frac{\partial}{\partial L_\lambda}
  K(\alpha_s,L_\lambda)
  = \left(
    \beta(\alpha_s)\frac{\partial}{\partial\alpha_s}-4\gamma_m(\alpha_s)
  \right)
  K(\alpha_s,L_\lambda)
\end{equation}
with $K(\alpha_s,L_\lambda)$ defined by
$\rho(\lambda)=(3/4\pi^2)\lambda^3 K(\alpha_s,L_\lambda)$.
By solving this equation at each order in $\alpha_s$, one can
determine the scale dependence of $\rho(\lambda)$, {\it i.e.}
the terms containing $L_\lambda$.
Since $\beta(\alpha_s)$ and $\gamma_m(\alpha_s)$ start at
$O(\alpha_s^2)$ and $O(\alpha_s)$ respectively, 
the $L_\lambda$ dependent term of
$\rho(\lambda)$ may be determined up to order $\alpha_s^{n+1}$
when a constant term, {\it i.e.} the coefficient at $L_\lambda=0$, is known at order $\alpha_s^n$.
The scale-dependent terms thus obtained are summarized in
Appendix~\ref{app:perturbative}.

We note that the relation
$\rho(\lambda)\propto\lambda^{4/(1+\gamma_m)-1}$,
suggested in studies of conformally invariant theories,
is valid in QCD only at one-loop level, {\it i.e.} 
from (\ref{eq:rhoMSbar}) and (\ref{eq:rho1}) one obtains
$\lambda^{3-8\alpha_s/\pi}$. This is consistent with the suggested
form $\lambda^{4/(1+\gamma_m)-1}$, with $\gamma_m=2\alpha_s/\pi$, at
one-loop level.
Beyond one-loop order, this correspondence does not hold.

We also introduce the exponent of the spectral function $F(\lambda)$,
defined by 
\begin{equation}
  \label{eq:F}
  F(\lambda) \equiv 
  \frac{\partial\ln\rho(\lambda)}{\partial\ln\lambda}.
\end{equation}
Its perturbative expansion is calculated 
through order $\alpha_s^4$
using the formula for $\rho(\lambda)$ summarized in Appendix~\ref{app:perturbative}:
\begin{equation}
  \label{eq:F^MS}
  F(\lambda) = 3 -
  F^{(1)}\frac{\alpha_s(\mu)}{\pi} -
  F^{(2)}\left(\frac{\alpha_s(\mu)}{\pi}\right)^2 -
  F^{(3)}\left(\frac{\alpha_s(\mu)}{\pi}\right)^3 -
  F^{(4)}\left(\frac{\alpha_s(\mu)}{\pi}\right)^4 
  + O(\alpha_s^5),
\end{equation}
where the coefficients $F^{(k)}$ for $N_f=3$ are 
\begin{eqnarray}
  \label{eq:F^(k)}
  F^{(1)} & = & 8,
  \\
  F^{(2)} & = & \frac{4}{3} \left(22 - 27 L_\lambda\right)
  \nonumber\\
          & = & 29.3333 - 36 L_\lambda,
  \\
  F^{(3)} & = & \frac{1}{36} \left( 6061 - 9216 L_\lambda + 
                5832 L_\lambda^2 - 1350 \pi^2 - 936 \zeta_3 \right)
  \nonumber\\
          & = & -233.003 - 256 L_\lambda + 162 L_\lambda^2,
  \\
  F^{(4)} & = & \frac{1}{5184} \left[
    \left( -3583861 + 1015200\pi^2 - 69984 \bar{\rho}^{(3)} + 3888\pi^4
           -315168\zeta_3 - 432000\zeta_5 \right)
    \right.
  \nonumber\\
    & & \left.
    +\left(-10980576 + 2624400\pi^2 + 1819584\zeta_3\right) L_\lambda
    + 8771328 L_\lambda^2 - 3779136 L_\lambda^3\right]
  \nonumber\\
          & = & -1348.6655 + 3300.2425 L_\lambda + 1692 L_\lambda^2 -
                729 L_\lambda^3.
\end{eqnarray}
Here, $\zeta_5\simeq$ 1.03692, and
$\bar{\rho}^{(3)}$ is the value of $\rho^{(3)}$ at $\lambda=\mu$,
which is given as
$-\bar{\rho}^{(3)}=-c_{43}/(64\pi^3)+c_{41}/(256\pi)$
using (\ref{eq:rho3}).
$F(\lambda)$ is defined in the $\overline{\mathrm{MS}}$ scheme at a
renormalization scale $\mu$,
and the coefficients $F^{(k)}$ have a dependence on
$L_\lambda=\lambda/\mu$. 
We may also obtain the term at $O(\alpha_s^5)$ except for the constant 
term, which comes from an as yet unknown constant $\bar{\rho}^{(4)}$ defined,
analogue with $\bar{\rho}^{(3)}$, as
\begin{eqnarray}
  \label{eq:F^(5)}
  F^{(5)} & = & \frac{1}{108864} \Bigl[
-30717197 + 85848525 \pi^2 + 6858432 \bar{\rho}^{(3)} - 
 16438275 \pi^4\nonumber\\
 &&-
  1959552 \bar{\rho}^{(4)} + 54000 \pi^6 - 
 51803465 \zeta_3 - 5443200 \pi^2 \zeta_3 \nonumber\\
 &&+
 73592064 \zeta_3 ^2 - 280468510 \zeta_5 + 
 179028360 \zeta_7
 \nonumber\\
 &&+
  L_\lambda\left(1526603652  - 395992800  \pi^2 + 26453952
    \bar{\rho}^{(3)}  \right.\nonumber\\
 && 
\ \ \ \ \ \ \ \left. -1469664  \pi^4 + 66824352  \zeta_3 + 163296000  \zeta_5\right)\nonumber\\
&&+
L_\lambda ^2\left(
1986475428 - 496011600 \pi^2 - 343901376 \zeta_3\right)\nonumber\\
&&- 1087551360 L_\lambda ^3 + 357128352 L_\lambda ^4
\Bigr]\nonumber\\
&=&
-18.0000 \bar{\rho}^{(4)} + 3751.43 
+24167.1 L_\lambda
-30518.4 L_\lambda ^2
-9990 L_\lambda^3
+\frac{6561}{2} L_\lambda ^4,
\nonumber\\
\end{eqnarray}
where $\zeta_7\simeq$ 1.00835. 

Let us attempt to estimate the size of the uncertainty in
$F(\lambda)$ due to the unknown constant $\bar{\rho}^{(4)}$.
As we will see below, the value of $\lambda$ we are going to use to
determine $\alpha_s(\mu)$ is in the range 0.8--1.2~GeV.
We therefore take $\lambda$ = 1.2~GeV as a representative value in the
following estimate. 
By setting $\mu=\lambda$, the series (\ref{eq:F^MS})
is simplified to
\begin{equation}
  \label{eq:Fmu}
  F(\lambda)_{\mu = \lambda} =
  3
  -2.54648\alpha_s
  -2.97209\alpha_s^2
  +7.51469\alpha_s^3
  +13.8454\alpha_s^4
  -d_5\alpha_s^5
  + \cdots,
\end{equation}
where the last term contains the unknown constant
$d_5=(-18.0000\bar{\rho}^{(4)}+3751.43)/\pi^5$.
Since the coefficients grow as fast as a factor of 2 from one
order to the next in (\ref{eq:Fmu}), it is natural to assume that the
size of $d_5$ is about 25.
To be conservative, we allow $\pm 50$ for the uncertainty in $d_5$.
Using $\alpha_s(\mu=1.2\mathrm{~GeV})$ = 0.406, 
the size of the $O(\alpha_s^5)$ term is then $\pm 0.55$, 
which is 18\% of the total.

If we choose a higher value for the renormalization scale, {\it e.g.}
$\mu=2.5\lambda$ or $5\lambda$, the series becomes
\begin{equation}
  \label{eq:Fmu_2.5}
  F(\lambda)_{\mu = 2.5\lambda} =
  3
  -2.54648\alpha_s
  -6.31432\alpha_s^2
  -4.43721\alpha_s^3
  +24.5484\alpha_s^4
  -(d_5-123.421)\alpha_s^5
  + \cdots,
\end{equation}
or
\begin{equation}
  \label{eq:Fmu_5}
  F(\lambda)_{\mu = 5\lambda} =
  3
  -2.54648\alpha_s
  -8.84261\alpha_s^2
  -19.3071\alpha_s^3
  -7.81967\alpha_s^4
  -(d_5-177.403)\alpha_s^5
  + \cdots,
\end{equation}
respectively.
We find that some coefficients are larger than those for $\mu=\lambda$
but not so much as to spoil the convergence. 
In fact, with $\alpha_s(\mathrm{3~GeV})$ = 0.244 and
$\alpha_s(\mathrm{6~GeV})$ = 0.191,
the size of the uncertainty from the unknown $O(\alpha_s^5)$ term is
1.4\% and 0.4\% for $\mu=2.5\lambda$ and $5\lambda$, respectively, if
we assume the same $\pm 50$ uncertainty for $d_5$.
Such reduction in uncertainty is of course achieved at the price of
potentially large higher order effects due to logarithmic
enhancements, {\it i.e.} $\sim\ln\lambda/\mu$, of the coefficients. 
We roughly estimate the growth of the $\mu$-dependent coefficient to
be a factor of 4 from one order to the next for the case of
$\mu=5\lambda$, 
and give an estimate $\sim\pm 700\alpha_s^6$ for the contribution of
the next order.
With $\alpha_s(\mathrm{6~GeV})$ = 0.204, this amounts to an uncertainty
of 2.6\% for $F(\lambda)$.
We use this estimate when we quote the value of
$\alpha_s(\mathrm{6~GeV})$ determined using (\ref{eq:Fmu_5}).

\begin{figure}[tbp]
  \includegraphics[width=13cm, angle=0]{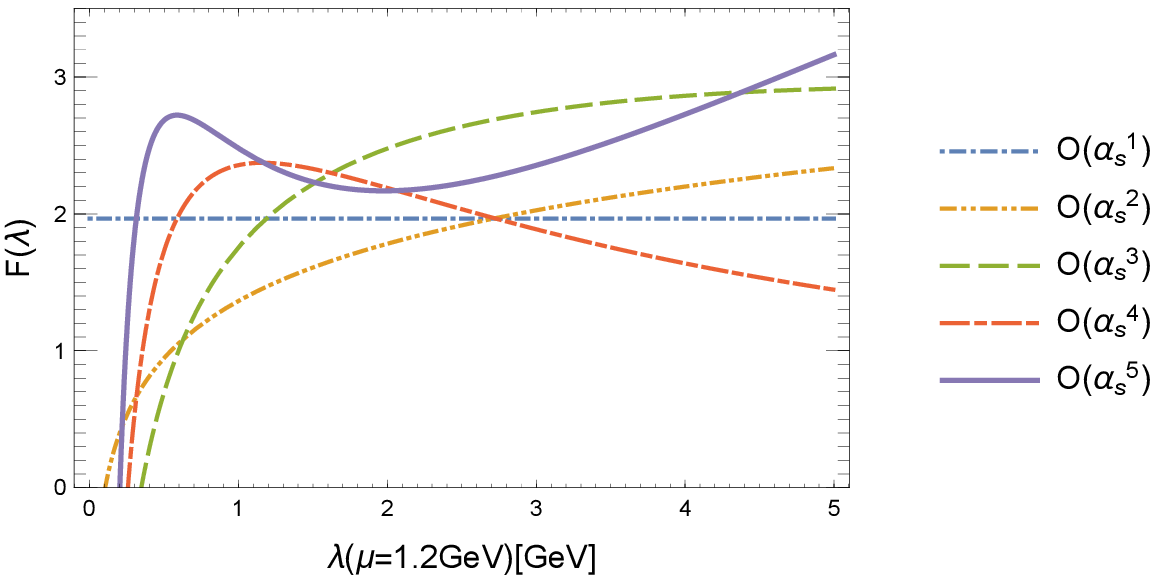}
  \includegraphics[width=13cm, angle=0]{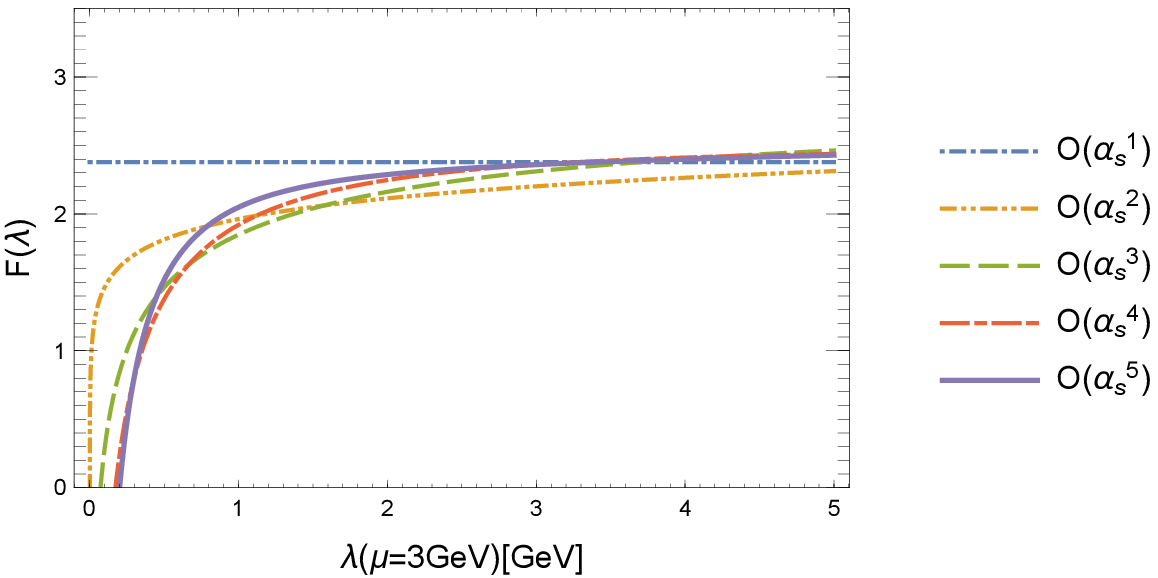}
  \includegraphics[width=13cm, angle=0]{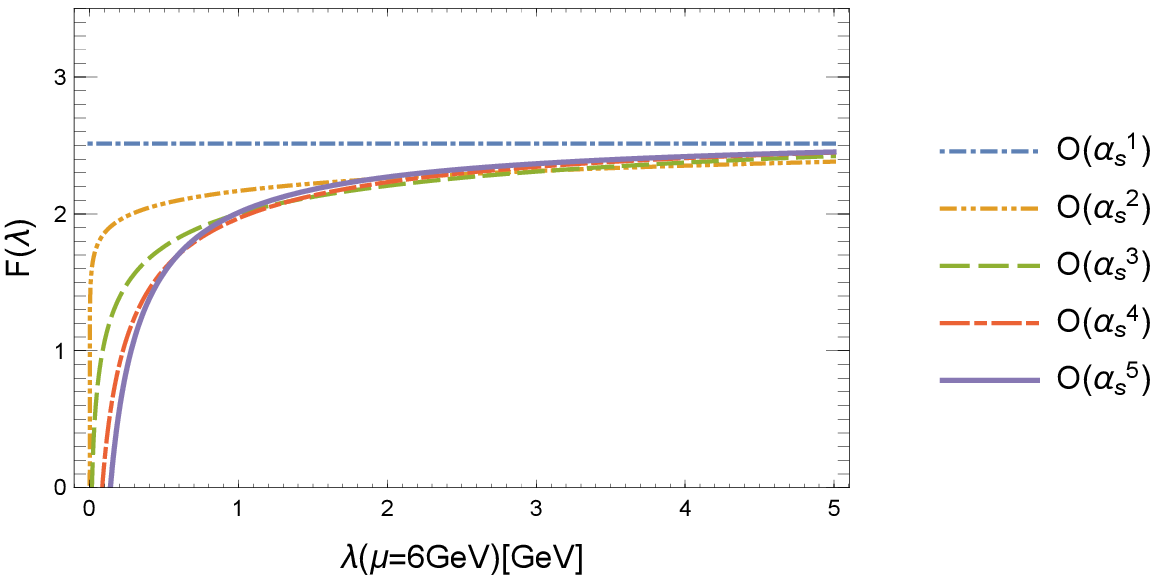}
  \caption{
    Convergence of the perturbative expansion for the exponent 
    $F(\lambda)$ of the Dirac spectral density in $N_f=3$ QCD.
    The Dirac eigenvalue is renormalized in the
    $\overline{\mathrm{MS}}$ scheme at the renormalization scale $\mu$
    = 1.2~GeV (upper panel), 3~GeV (middle) and 6~GeV (lower panel).
    The perturbative series is truncated at different orders from
    $O(\alpha_s)$ to $O(\alpha_s^5)$.
    The unknown constant term $d_5$ is set to zero.
  }
  \label{fig:pert_each}
\end{figure}

Figure~\ref{fig:pert_each} shows the convergence of the perturbative
series for $F(\lambda)$.
The renormalization scale $\mu$ is set to 1.2~GeV (upper panel), 3~GeV
(middle panel) and 6~GeV (lower panel), 
and the unknown higher order constant $d_5$ is set to zero.
It is clear that the choice of the renormalization scale 
$\mu$ = 1.2~GeV gives a bad convergence everywhere in $\lambda$.
On the other hand, we find the series converges well when $\mu$ is
taken to be 3~GeV or higher.
In particular, the convergence for $\lambda$ = 3~GeV or larger is
reasonably good for both $\mu=2.5\lambda$ and $5\lambda$.
The range of convergence extends to lower values of $\lambda$, down
to $\approx$ 1~GeV, when $\mu$ is set to the {\it higher} value, 
{\it i.e.} 6~GeV.
This is slightly counter-intuitive because it is generally recommended that the renormalization scale
$\mu$ is set to be close to the scale of interest,
which in this case is $\mu\approx\lambda$.
The optimal scale does, however, depend on the quantity; for
$F(\lambda)$ it turns out that the higher renormalization scale
yields better convergence.

Below $\lambda$ = 0.8~GeV, the convergence gets worse even with $\mu$
= 6~GeV,
{\it i.e.} the difference between third and higher orders becomes
more significant.
The main lattice results we use in this analysis are at $\lambda$ =
1.2~GeV or slightly lower, so the choice of $\mu$ = 6~GeV has the
advantage of better convergence.

\section{Discretization effects with lattice domain-wall fermion}
\label{sec:Tree_level}
Since we are interested in the relatively high energy region ($\sim$
1~GeV or higher) of the Dirac eigenvalue spectrum, 
discretization effects could be sizable on the lattices with inverse
lattice spacing $1/a=$ 2.4--4.6~GeV.
In this section we examine the size of lattice artifacts
that affects the Dirac spectrum using the free field limit.
The lattice artifacts depend on the details of the lattice formulation.
We consider here the domain-wall fermion formulation 
\cite{Kaplan:1992bt,Shamir:1993zy}, 
which we used in the numerical simulations.

The domain-wall fermion and its M\"obius generalization
\cite{Brower:2012vk} are formulated on a five-dimensional (5D) Euclidean 
space, and the mapping to four-dimensional (4D) space is given by
\begin{equation}
  \label{eq:5to4}
  aD_{\mathrm{ov}}(m_f,m_{\mathrm{PV}})\equiv 
  \left(2-(b-c)M_0\right) M_0am_{\mathrm{PV}}
  \left[
    {\cal P}^{-1} D_{\mathrm{DW}}^{-1}(m_{\mathrm{PV}}) D_{\mathrm{DW}}(m_f) 
    {\cal P}
  \right]_{11}.
\end{equation}
Here, $D_{\mathrm{DW}}(m_f)$ denotes the 5D Dirac operator of the
domain-wall fermion with a fermion mass $m_f$.
The dimensionless parameters $b$, $c$ and $M_0$ control the kernel
operator as described below.
The 5D operator $D_{\mathrm{DW}}(m_f)$ is multiplied by the inverse of
another 5D Dirac operator with a heavier mass $m_{\mathrm{PV}}$, which 
plays the role of Pauli-Villars
regulator that cancels the bulk mode in the 5D space.
The Pauli-Villars mass $m_{\mathrm{PV}}$ is usually taken to be $1/a$.
The 5D operators are then sandwiched by a permutation operator 
${\cal P}$ and its inverse to bring the relevant 4D degrees of freedom
to a 4D surface of the 5D space, and its surface component ``11'' is
taken. 
(See \cite{Brower:2012vk,Boyle:2015vda} for details and further
references.) 
Equation (\ref{eq:5to4}) leads to an overlap-Dirac operator
of the form
\begin{equation}
  \label{eq:Dov}
  aD_{\mathrm{ov}}(m_f,m_{\mathrm{PV}}) =
  (2-(b-c)M_0)M_0 am_{\mathrm{PV}}
  \frac{(1+am_f)+(1-am_f)\gamma_5 \mathrm{sgn}(\gamma_5 aD_M)}{
        (1+am_{\mathrm{PV}})+(1-am_{\mathrm{PV}})
        \gamma_5 \mathrm{sgn}(\gamma_5 aD_M)}
\end{equation}
in the limit of large fifth dimension $L_s\to\infty$.
The overlap operator \cite{Neuberger:1998wv} is realized with a polar
approximation to the sign function ``sgn''.
Note that the conventional form is recovered for $am_{\mathrm{PV}}=1$.
The M\"obius kernel operator $D_M$ is given by
\begin{equation}
  \label{eq:kernel}
  aD_M = \frac{(b+c) aD_W}{2 + (b-c) aD_W}
\end{equation}
in terms of the Wilson-Dirac operator $D_W$.
The mass parameter $M_0$ is hidden in the definition of $D_W$ as a
large negative mass term of order $1$ in lattice units.
We take $M_0=1$ in this work.
The parameters $b$ and $c$ determine the kernel $D_M$.
$b-c=1$ corresponds to the original domain-wall fermion, 
which we use in this work.

The general form of the 4D overlap operator (\ref{eq:Dov}) satisfies
the Ginsparg-Wilson relation 
\begin{equation}
  D_{\mathrm{ov}}^{-1}(0,m_{\mathrm{PV}}) \gamma_5 +
  \gamma_5 D_{\mathrm{ov}}^{-1}(0,m_{\mathrm{PV}})
  =
  \frac{2a}{(2-(b-c)M_0)M_0am_{\mathrm{PV}}}\gamma_5.
\end{equation}
It can also be shown that the exponential locality of the 4D effective
operator is satisfied for any non-zero value of $am_{\mathrm{PV}}$.
(See Appendix~\ref{sec:locality_and_chirality} for details.)
Therefore, domain-wall fermions with any $am_{\mathrm{PV}}\neq 0$ are
a theoretically valid choice.
In this work we use the standard choice $am_{\mathrm{PV}}=1$ in the
numerical simulation of sea quarks, 
and reinterpret the results to the cases of $am_{\mathrm{PV}}\neq 1$
for valence quarks.

We calculate the eigenvalues of the Hermitian operator
$a^2D_{\mathrm{ov}}^\dagger(m_f,m_{\mathrm{PV}})D_{\mathrm{ov}}(m_f,m_{\mathrm{PV}})$
in the massless limit, $m_f=0$.
Leaving the dependence on $am_{\mathrm{PV}}$, we denote its eigenvalue as
$a^2\lambda^2(am_{\mathrm{PV}})$.
Since $a^2D_{\mathrm{ov}}^\dagger D_{\mathrm{ov}}$'s with different values of
$am_{\mathrm{PV}}$ commute with each other, their eigenvalues are
simply related by 
\begin{equation}
  a^2\lambda^2(am_{\mathrm{PV}})
  = \frac{a^2\lambda^2(1)}{\displaystyle
    1 + \left(\frac{1}{a^2m_{\mathrm{PV}}^2} - 1\right) 
    \frac{a^2\lambda^2(1)}{(2-(b-c)M_0)^2M_0 ^2}}.
\label{eq:gpv_rel}
\end{equation}
In the limit $am_{\mathrm{PV}}\to\infty$ this corresponds to the
projection of the eigenvalue of $aD_{\mathrm{ov}}$ in the complex
plane to the imaginary axis, which was adopted in
\cite{Fukaya:2009fh,Fukaya:2010na,Cossu:2016eqs}.

At tree-level, it is straightforward to calculate the eigenvalue 
$a^2\lambda^2$ for a plane wave state with a given four-momentum $ap_\mu$. 
To obtain the spectral density, we count the number of states that
give an eigenvalue in an interval $[a^2\lambda^2,a^2(\lambda\pm\delta\lambda)^2]$
from randomly chosen points of momenta $ap_\mu$, each component of
which is in the range $ap_\mu=[-\pi,+\pi]$.
We generate a large number of points (up to $10^{12}$) so that the
statistical error becomes negligible for our choice of bin size
$a\delta\lambda=0.0025$ for the computation of $\rho(\lambda)$.


\begin{figure}[tbp]
  \includegraphics[width=9cm,angle=-90]{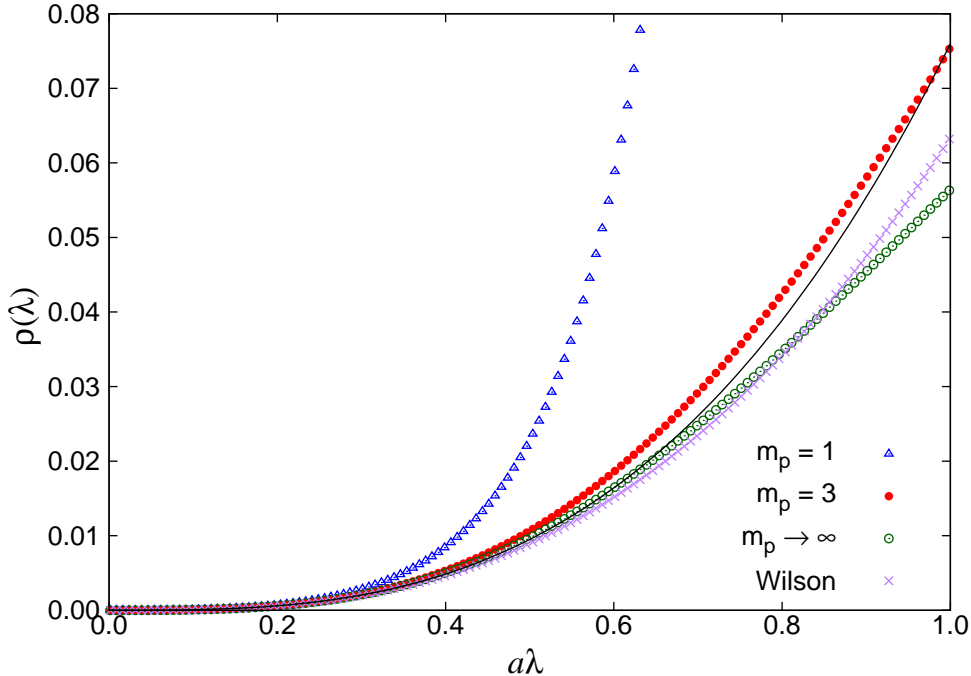}
  \caption{
    Dirac spectral density $\rho(\lambda)$ at tree-level.
    The results for domain-wall fermions with
    $am_{\mathrm{PV}}=1$ (triangles), $am_{\mathrm{PV}}=3$ (filled circles),
    $am_{\mathrm{PV}}\to\infty$ (open circles), and for 
    Wilson fermion (crosses) are shown.
    The black solid line represents the continuum result, 
    $\rho(\lambda) = 3\lambda ^3/4\pi^2$.
  }
\label{fig:tree_spectral}
\end{figure}

The results for domain-wall fermions with various $am_{\mathrm{PV}}$ 
as well as that for Wilson fermion are shown in
Figure~\ref{fig:tree_spectral}.
All formulations coincide with the continuum limit
$\rho(\lambda) = 3\lambda ^3/4\pi^2$
below $a\lambda\simeq$ 0.3, as expected.
Above this value, on the other hand, the spectral density for the
domain-wall fermion with Pauli-Villars mass $am_{\mathrm{PV}}=1$
overshoots the continuum curve very rapidly.
This is understood to be because the maximum eigenvalue with
$am_{\mathrm{PV}}=1$ is $a\lambda_{\mathrm{max}}(am_{\mathrm{PV}})=1$, 
so high eigenvalues tend to rapidly increase in frequency $a\lambda$ increases. 
With $am_{\mathrm{PV}}>1$, the maximum eigenvalue is stretched
by a factor of $am_{\mathrm{PV}}$ (when $b-c=M_0=1$) as indicated by
(\ref{eq:gpv_rel}). 
This allows the spectral densities for
$am_{\mathrm{PV}}=3$ and $\infty$ follow that of the continuum theory
up to $a\lambda\sim$ 0.5--0.6.
The spectral density for Wilson fermions also resembles the continuum
spectrum in the same region. 

We now consider the exponent $F(\lambda)$ of the spectral density
(\ref{eq:F}). 
We approximate the differential $\partial/\partial\lambda$
by a symmetric difference:
\begin{equation}
  \frac{\partial\rho}{\partial\lambda}
  \simeq \frac{\rho(\lambda+\delta\lambda)-\rho(\lambda-\delta\lambda)}{
    2\delta\lambda},
  \label{eq:diff}
\end{equation}
which is valid up to an error of $O(\delta\lambda^2)$.
For the continuum spectrum $\rho(\lambda)=3\lambda^3/4\pi^2$,
the leading relative error for $F(\lambda)$ is given by 
$2\delta\lambda^2/\lambda^2$.
In order that this source of error is below 0.01, 
$\delta\lambda/\lambda<0.07$ has to be satisfied.
With our choice of $a\delta\lambda$ = 0.005, adopted for the study of
$F(\lambda)$, the reliable range of the approximation is
$a\lambda>0.07$. 

\begin{figure}[tbp]
  \includegraphics[width=9cm, angle=-90]{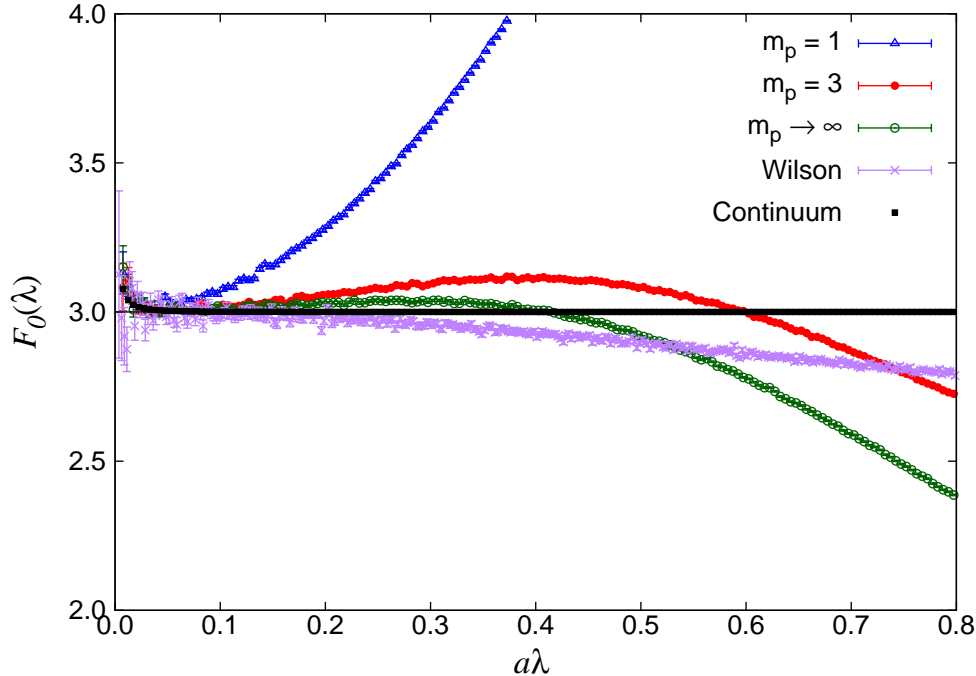}
  \caption{
    Exponent of the Dirac spectral density
    $F(\lambda)=\partial\rho(\lambda)/\partial\lambda$
    evaluated at tree-level for various choices of the lattice
    fermion formulation.
    The standard domain-wall fermion, corresponding to
    $am_{\mathrm{PV}}=1$, is represented by triangles, 
    and its modifications with different values of the
    Pauli-Villars mass are given by filled circles ($am_{\mathrm{PV}}=3$)
    and open circles ($am_{\mathrm{PV}}\to\infty$). 
    The results for the Wilson fermion is also plotted (crosses).
    The derivative of the spectral density is evaluated numerically
    for each bin of size $a\delta\lambda=0.005$.
    For comparison, the numerical derivative of the free field spectral
    density, $\sim\lambda^3$, is represented by black dots, which indicate
    the size of the error due to finite bin size.
  }
  \label{fig:tree_ano_dim}
\end{figure}

Figure~\ref{fig:tree_ano_dim} shows $F_0(\lambda)$,
which is $F(\lambda)$ evaluated at the tree level, with bin size
$a\delta\lambda=0.005$.
It is compared with the corresponding continuum result, which is
a constant, {\it i.e.} 3.
In the plot, the continuum theory is also calculated with the same
finite difference (\ref{eq:diff}) (black dots).
Some deviation from 3 can be seen near $a\lambda=0$, which indicates
the size of the systematic error due to the discretized derivative.
Our purpose is to extract the exponent in the high energy region, so we 
ignore the points below $a\lambda<$ 0.025, thus this source of error
is negligible.

The discretization effects we find for the spectral density
(Figure~\ref{fig:tree_spectral}) are magnified in its derivative as
one can see in Figure~\ref{fig:tree_ano_dim}. 
The standard domain-wall fermion, {\it i.e.} with
$am_{\mathrm{PV}}=1$, already shows a significant deviation from the continuum
theory below $a\lambda\sim$ 0.1.
That is improved with the other choices: 
$am_{\mathrm{PV}}=3$ or $\infty$. 
In order to parametrize the discretization effect, 
we approximate the curve by a polynomial in $(a\lambda)^2$ and
numerically obtain
\begin{align}
  F_0(\lambda) & = 3 + 6.93 (a\lambda)^2 + 2.08 (a\lambda)^4 
  -13.0(a\lambda)^6 + 24.9(a\lambda)^8
             & (am_{\mathrm{PV}}=1),
  \nonumber\\
  \label{eq:F_tree}
  F_0(\lambda) & = 3 + 1.59 (a\lambda)^2 - 6.58 (a\lambda)^4
  + 6.95 (a\lambda)^6 - 2.51 (a\lambda)^8 
             & (am_{\mathrm{PV}}=3),
\end{align}
which reproduce the data well for $a\lambda<$ 0.6.
The smaller coefficients for $am_{\mathrm{PV}}=3$ confirm the
suppression of the discretization effect.



In the following analysis we choose $am_{\mathrm{PV}}=3$.
Namely, we convert the eigenvalues from
$\lambda(am_{\mathrm{PV}}=1)$ to $\lambda(am_{\mathrm{PV}}=3)$ using (\ref{eq:gpv_rel}).
We then use the parametrization (\ref{eq:F_tree}) to further correct
the lattice data by multiplying by $3/F_0(\lambda)$
to eliminate the leading discretization effects.
Finally we take the continuum limit of the lattice data
assuming that the remaining effects are linear in $a^2$.
This assumption is justified by inspecting the real data.

Strictly speaking, the fermion formulation is different between
sea and valence when we choose different values of $am_{\mathrm{PV}}$, 
and the action of the whole system represents that of
a partially quenched theory.
However, since the correspondence between the eigenvalues of each
formulation is one-to-one even in the interacting case, 
the continuum limit is guaranteed to be the same.
It can be seen explicitely in the relations (\ref{eq:gpv_rel}), as 
$a^2\lambda^2(am_\mathrm{PV}) = a^2\lambda^2(1) + O(a^4\lambda^4(1))$.
We focus on the region small $a\lambda(1)$ for which the results from
$a\lambda(am_\mathrm{PV})$
are reliable in the continuum limit.

Another concern may be about the validity of the formulation with
$am_{\mathrm{PV}}\neq 1$.
As discussed in Appendix~\ref{sec:locality_and_chirality} the
localization length for the 4D effective operator (\ref{eq:5to4})
stays finite for finite $am_{\mathrm{PV}}$.
Our choice, $am_{\mathrm{PV}}=3$, keeps the localization length about
the same as that of $am_{\mathrm{PV}}=1$, and the formulation is valid
with respect to the locality property.

In Figure~\ref{fig:tree_ano_dim} we also plot the result with the
standard Wilson fermion (crosses), which shows a smooth and mild
continuum limit. 
Its polynomial approximation gives
\begin{align}
  F_0(\lambda) & = 3 - 0.459 (a\lambda)^2 + 0.208 (a\lambda)^4 
             & (\mathrm{Wilson}),
\end{align}
and the size of the $O(a^2)$ terms is even smaller than that for the
domain-wall fermion $am_{\mathrm{PV}}=3$ or $\infty$.

\section{Lattice calculation}
\label{sec:lattice_details}

\subsection{Lattice ensembles}

\begin{table}
\begin{tabular}{cccc|cc|cc} 
  \hline\hline
  $\beta$ & $a^{-1}$ &
  $L^3\times T(\times L_s)$ &
  \#meas &
  $am_{ud}$ & $am_{s}$  & 
  $m_{\pi}$ & $m_\pi L$ 
  \\
  & [GeV] & & & & & [MeV] & 
  \\ \hline
  4.17 & 2.453(4) &$32^3\times 64(\times 12)$ & 100& 
  0.0035 & 0.040  & 230(1) & 3.0 \\ 
  &&&&
  0.007    & 0.030  & 310(1) & 4.0 \\
  &&&&
  0.007 & 0.040  & 309(1) & 4.0 \\
  &&&&
  0.012    & 0.030  & 397(1) & 5.2 \\
  &&&&
  0.012    & 0.040     & 399(1) & 5.2 \\
  &&&&
  0.019    & 0.030     & 498(1) & 6.5 \\
  &&&&
  0.019    & 0.040  & 499(1) & 6.5 \\
  &&
  $48^3 \times 96(\times 12)$ & 100 &
  0.0035 & 0.040  & 226(1) & 4.4 \\
  \hline 
  4.35 & 3.610(9) & $48^3\times 96(\times 8)$ & 50 &
  0.0042 & 0.0180  & 296(1) & 3.9 \\ 
  &&&&
  0.0042 & 0.0250  & 300(1) & 3.9 \\
  &&&&
  0.0080 & 0.0180  & 407(1) & 5.4 \\ 
  &&&&
  0.0080 & 0.0250  & 408(1) & 5.4 \\ 
  &&&&
  0.0120 & 0.0180  & 499(1) & 6.6 \\ 
  &&&&
  0.0120 & 0.0250  & 501(1) & 6.6 \\
  \hline
  4.47 & 4.496(9) & $64^3 \times 128(\times 8)$ & 39 &
  0.0030 & 0.015  & 284(1) & 4.0 \\
  \hline\hline
\end{tabular}
\caption{
  Lattice ensembles used in this study.
}
\label{tab:setup}
\end{table}

Our lattice QCD simulations are performed with $2+1$ flavors of
dynamical quarks. 
We use the M\"obius domain-wall fermion \cite{Brower:2012vk}
for the fermion formulation.
We use a tree-level Symanzik improved action is used as the gauge action,
and stout smearing \cite{Morningstar:2003gk} is applied to the
gauge link when coupled to the fermions.
With this setup, the Ginsparg-Wilson relation is well satisfied,
{\it i.e.}
the residual mass as an indicator of chiral symmetry violation is of  
$O(\mbox{1~MeV})$ on our coarsest lattice and even smaller on the finer
lattices. 

We use 15 gauge ensembles with different parameters as listed in Table~\ref{tab:setup}.
Lattice spacings are $a$ = 0.080, 0.055, and 0.044~fm.
The size of these lattices ($L/a$ = 32, 48, and 64) is chosen such
that the physical size $L$ is kept constant: 2.6--2.8~fm.
The temporal size $T/a$ is always twice as long as $L/a$.
Degenerate up and down quark masses $am_{ud}$ cover a range of pion
masses $m_\pi$ between 230 and 500~MeV.
A measure of the finite volume effect $m_\pi L$, is also listed in
the table, but finite volume effects are irrelevant for this
work as we are interested in high-energy observables.
The strange quark mass $am_s$ is taken to be close to its physical value.
Sea quark mass dependence of the spectral density turns out to be
negligible in the high-energy region.
For valence quarks, we only use the massless Dirac operator to count the number of eigenvalues.

We accumulate 10,000 molecular dynamics trajectories, from which we
choose 50--100 equally separated gauge configurations for the
calculation of the eigenvalue spectrum.
The number of measurement, ``\#meas'' in the table,
is 39--100 depending on the ensemble.

The lattice scale is determined through the Wilson-flow time $t_0$
\cite{Luscher:2010iy} with an input
$t_0^{1/2}$ = 0.1465(21)(13)~fm
\cite{Borsanyi:2012zs}.
The resulting values of $a^{-1}$ in the chiral limit
are listed in Table~\ref{tab:setup}, where the error given for
$a^{-1}$ is from the statistical error in the evaluation of the
Wilson flow time.
The error from the input value is taken into account separately.

Details of the ensemble generation are available in \cite{Kaneko:2013jla,Noaki:2014ura}.
The same gauge ensembles have so far been used for a calculation of
the $\eta'$ meson mass \cite{Fukaya:2015ara}, 
analysis of short-distance current correlators
\cite{Tomii:2015exs,Tomii:2016xiv,Tomii:2017cbt}, 
a determination of the charm quark mass from charmonium correlators
\cite{Nakayama:2016atf}, as well as
calculations of heavy-light meson decay constants
\cite{Fahy:2015xka}, 
and semi-leptonic decays
\cite{Kaneko:2017sct,Colquhoun:2017gfi,Hashimoto:2017wqo}.
We use the IroIro++ code set for lattice QCD \cite{Cossu:2013ola}.

For the renormalization of the Dirac eigenvalue, we multiply by the
renormalization constant 
$Z_m(\mathrm{2~GeV})=Z_S^{-1}(\mathrm{2~GeV})$ to obtain the value
defined in the $\overline{\mathrm{MS}}$ scheme:
\begin{equation}
  \lambda^{\overline{\mathrm{MS}}}(\mathrm{2~GeV})
  = Z_m(2\mathrm{~GeV})
  \lambda.
\end{equation}
The renormalization factor was determined in the analysis of the 
short-distance current correlator of light quarks using
continuum perturbation theory at $O(\alpha_s^4)$ 
\cite{Tomii:2016xiv}. 
The numerical values are $Z_S(\mathrm{2~GeV})=$
1.0372(145) at $\beta$ = 4.17, 
0.9342(87) at $\beta$ = 4.35, and 
0.8926(67) at $\beta$ = 4.47,
where the errors include statistical and systematic errors added
in quadrature. 
The spectral function is also renormalized in the
$\overline{\mathrm{MS}}$ scheme using the same renormalization
constant $Z_S(\mathrm{2~GeV})$.
This, however, is implicitly done in our calculation when we calculate
the eigenvalue density by dividing the number of eigenvalues in a
given bin of $\lambda$ by the corresponding bin size.

When we compare the results with perturbation theory,
we evolve the scale at which we renormalize the eigenvalue from 2~GeV to
6~GeV in order to use the more convergent perturbative series as 
discussed in Section~\ref{sec:Tree_level}.

\subsection{Stochastic calculation of the spectral density}
On each ensemble, we calculate the eigenvalue density of the Dirac
operator by using a stochastic estimator.
We outline the method in this section.
The details are available in \cite{Cossu:2016eqs}.

Suppose we are able to construct a filtering function $h(x)$ that
gives a constant (=1) only in a given range $[v,w]$ and is zero
elsewhere. 
The number of eigenvalues of a hermitian matrix $A$ in that range can 
then be estimated stochastically as
\begin{equation}
  n[v,w]\simeq \frac{1}{N}\sum_{k=1}^N \langle\xi_k h(A)\xi_k\rangle,
\end{equation}
where $N$ is the
number of normalized random noise vectors, $\xi_k$.
An approximation of the filtering function can be constructed using
the Chebyshev polynomial $T_j(x)$ through
\begin{equation}
  h(x) \simeq \sum^p _{j=0} \gamma_j T_j(x)
\end{equation}
with numerical coefficients $\gamma_j$, which are calculable as
functions of $v$ and $w$.
The approximation is valid in the range $-1\leq x\leq 1$.
In practice, one also introduces a stabilization factor $g_j^p$ 
in addition to $\gamma_j$. 
The Chebyshev polynomial of a given matrix $A$ can be easily
calculated using a recursion relation:
$T_0(x)=1$, $T_1(x)=x$, and $T_j(x)=2xT_{j-1}(x)-T_{j-2}(x)$.
We constructed the polynomial up to order $p$ = 8,000--16,000
depending on the ensembles. 
Details of the method are found in \cite{Cossu:2016eqs,Napoli}.
See also \cite{Fodor:2016hke} for a related work.

The mode number $n[v,w]$ can then be calculated by
\begin{equation}
  n[v,w] \simeq \frac{1}{N}\sum_{k=1}^N \sum_{j=0}^p g_j^p\gamma_j
  \langle\xi_k^\dagger T_j(A)\xi_k\rangle.
\end{equation}
An important point is that the computationally intensive part
$\langle\xi_k^\dagger T_j(A)\xi_k\rangle$ is independent of the range
$[v,w]$.
Once the inner products
$\langle\xi_k^\dagger T_j(A)\xi_k\rangle$ are calculated for all
$j$ below some upper limit $p$, they can be combined with the known
factors $g_j^p\gamma_j$ to construct the estimate for $n[v,w]$ for
arbitrary $[v,w]$.
We use this property to obtain the whole eigenvalue spectrum from
$\lambda=0$ to the upper limit at the order of the lattice cutoff.

We apply this technique to a hermitian operator
$A=2a^2D_{\mathrm{ov}}(m_f,1)^\dagger D_{\mathrm{ov}}(m_f,1)-1$,
whose range is $[-1,+1]$.
The spectral function $\rho (\lambda)$ can be obtained from the mode 
number $n[v,w]$ with
$v=2a^2(\lambda-\delta/2)^2-1$,
$w=2a^2(\lambda+\delta/2)^2-1$.
Here we introduce a bin size $\delta$.
We then obtain the estimate of the spectral function at $\lambda$ by
\begin{equation}
  a^3\rho(\lambda)
  =
  \frac{1}{2V/a^4}\frac{n[v,w]}{a\delta}.
\end{equation}
The error due to the finite bin size is taken into account when we
calculate the derivative of $\rho(\lambda)$.
The systematic effect due to the truncation of the Chebyshev
polynomial grow for smaller bin sizes.
In our calculation, we confirm that this error is less than the
statistical error for our choice of bin size.

\section{Lattice results for the spectral function and its exponent}
\label{sec:eigen_lat}

Following the method outlined in the previous section, 
we calculate the spectral density $\rho(\lambda)$ on all ensembles in Table~\ref{tab:setup}.
We introduce one noise vector per configuration.
We present the results in the $\overline{\mathrm{MS}}$ scheme.
The renormalization scale $\mu$ is set to 2~GeV for the
spectral density, and is transformed to 6~GeV when we discuss the 
matching of the exponent $F(\lambda)$ with perturbation theory. 
The bin size $\delta$ = 0.05~GeV is chosen in physical units
(for the renormalization scale of 2~GeV)
such that the error due to the truncation of the Chebyshev polynomial 
is less than the statistical error for any single bin used in the
analysis. 

\begin{figure}[tbp]
  \centering
  \includegraphics[width=6.5cm, angle=-90]{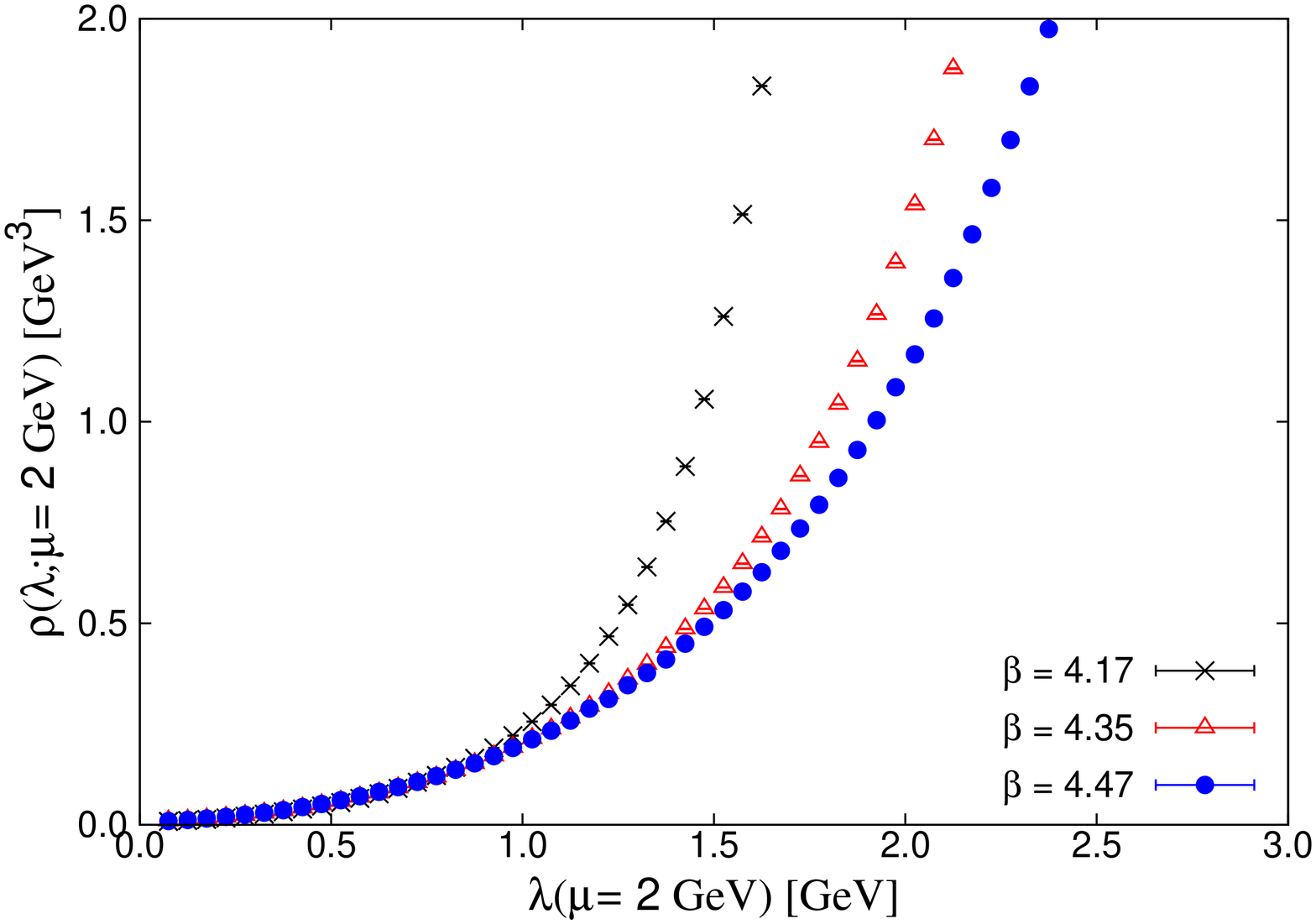}
  \includegraphics[width=6.5cm, angle=-90]{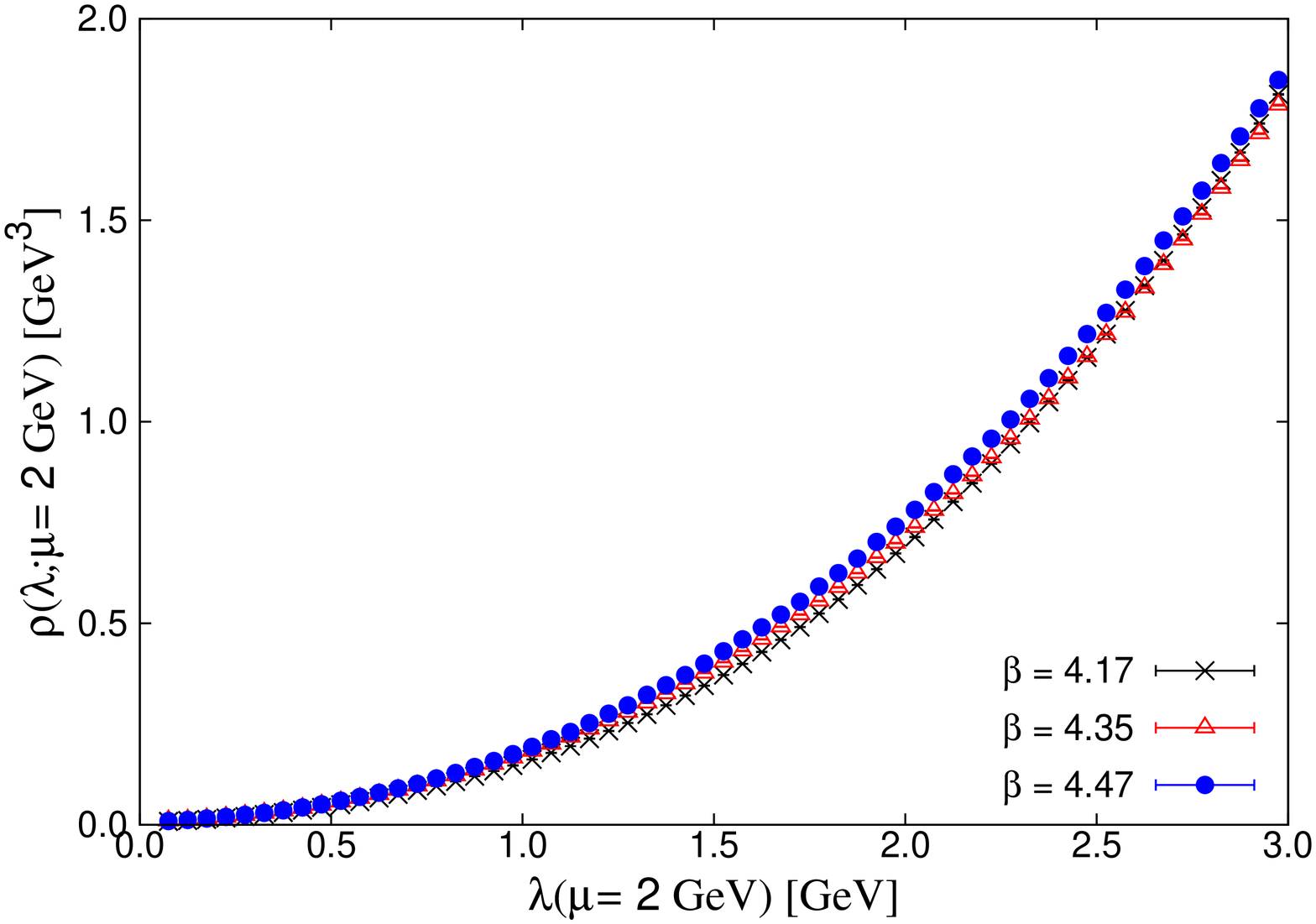}
  \includegraphics[width=6.5cm, angle=-90]{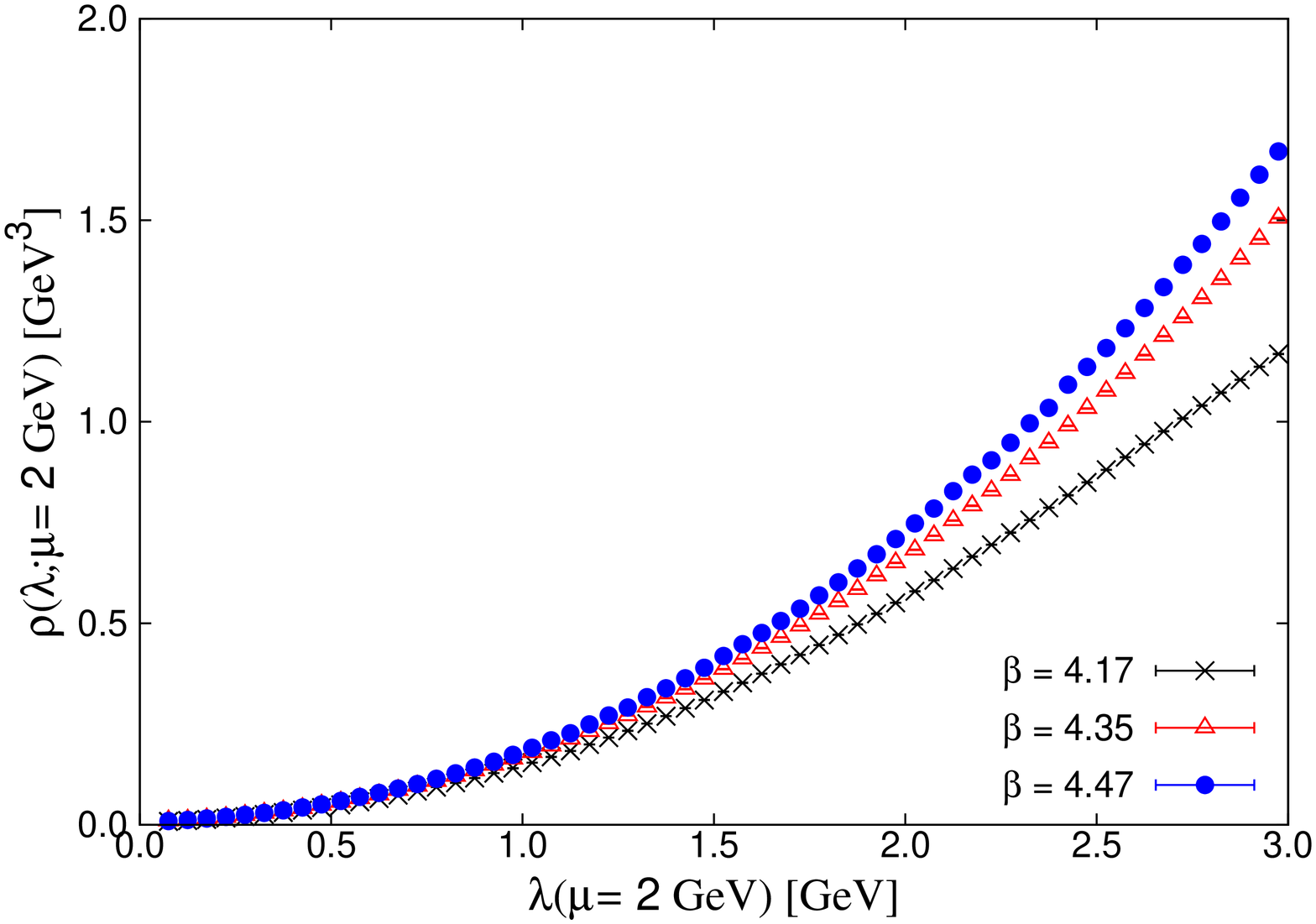}
  \caption{
    Dirac spectral density $\rho(\lambda)$ as a function of
    $\lambda(\mu=2\mathrm{~GeV})$ with Pauli-Villars mass 
    $am_{\mathrm{PV}}=1$ (top panel), $am_{\mathrm{PV}}=3$ (middle),
    and $am_{\mathrm{PV}}\to\infty$ (bottom). 
    We plot the data at $\beta=4.17$ (crosses),
    $\beta=4.35$ (triangles) and $\beta=4.47$ (circles).
    Each set of points is obtained after averaging over ensembles with
    different sea quark masses.
  }
  \label{fig:simu_spectral}
\end{figure}

Figure \ref{fig:simu_spectral} shows the Dirac spectral density
calculated at each $\beta$ value with different Pauli-Villars
masses: $am_{\mathrm{PV}}$ = 1 (top panel), 3 (middle), and $\infty$
(bottom panel).
Lattice data are averaged over ensembles with different sea quark
masses, since the dependence on the sea quark mass is negligible
except in the region near $\lambda\approx 0$.
We see only a tiny dependence on the lattice spacing $a$ or on the 
Pauli-Villars mass $am_{\mathrm{PV}}$ in the low-lying Dirac
eigenvalue spectrum below, say, $\lambda$ = 0.5~GeV.
On the other hand, relatively higher Dirac eigenvalues depend
significantly on the lattice spacing, especially for
$am_{\mathrm{PV}}=1$. 
For the choices $am_{\mathrm{PV}}=3$ and $am_{\mathrm{PV}}\to\infty$,
the scaling violation is milder than anticipated from the tree-level
analysis discussed in Section~\ref{sec:Tree_level}.

As described in Sections \ref{sec:Ano_dim} and \ref{sec:Tree_level},
we extract the exponent $F(\lambda)$ of the Dirac eigenvalue spectral 
density $\rho(\lambda)$. 
The derivative in terms of $\lambda$ is numerically performed using 
(\ref{eq:diff}).
For bin size $\delta\lambda$ = 0.05~GeV, the systematic error due 
to the discretized derivative is about 0.008 (0.3\%) in the
lowest bin used in the final analysis, 
$\lambda_{\mathrm{min}}$ = 0.8~GeV. 
This is smaller than the statistical error of the corresponding data
point. 

\begin{figure}[tbp]
  \centering
  \includegraphics[width=6.5cm,angle=-90]{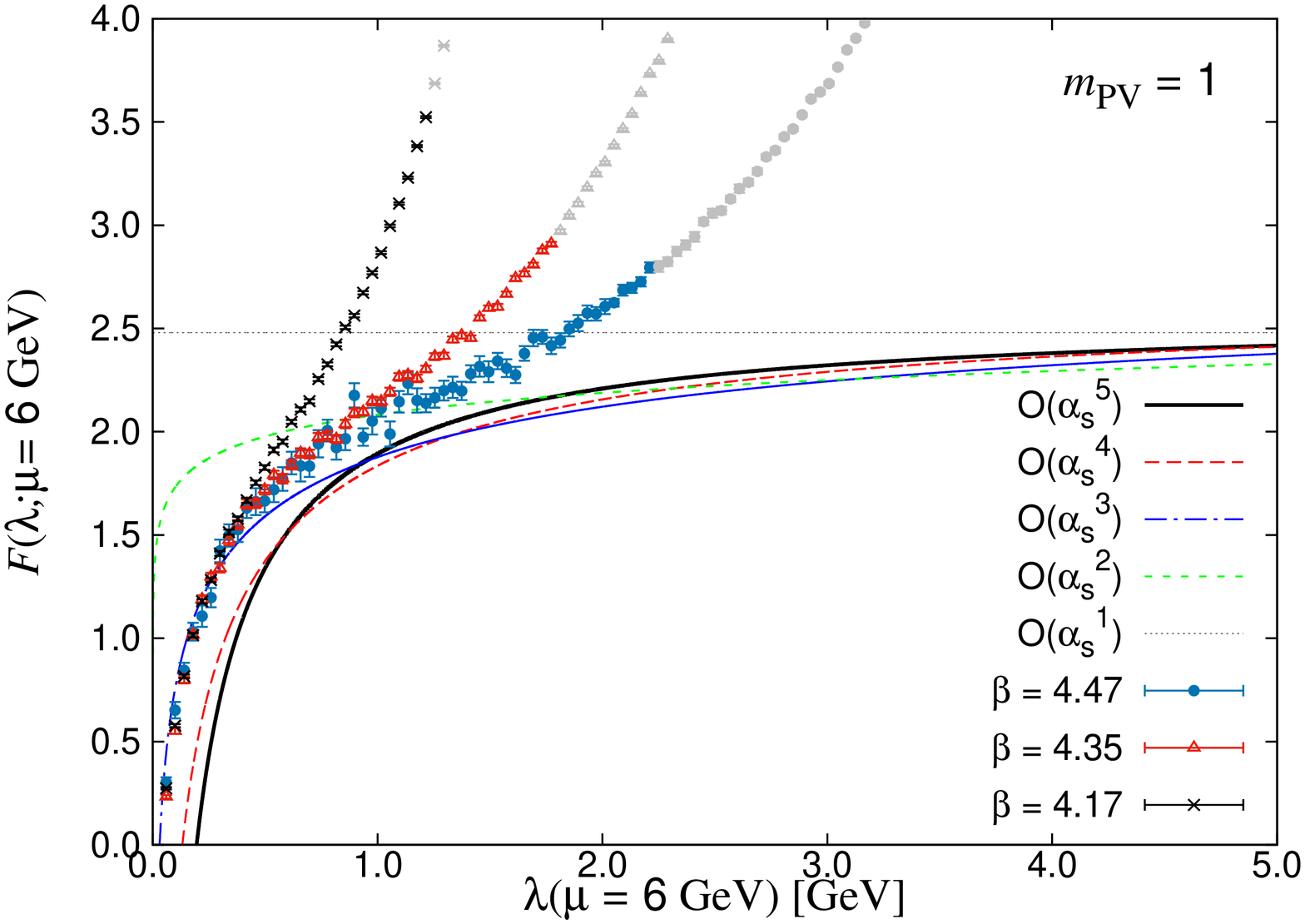}
  \includegraphics[width=6.5cm,angle=-90]{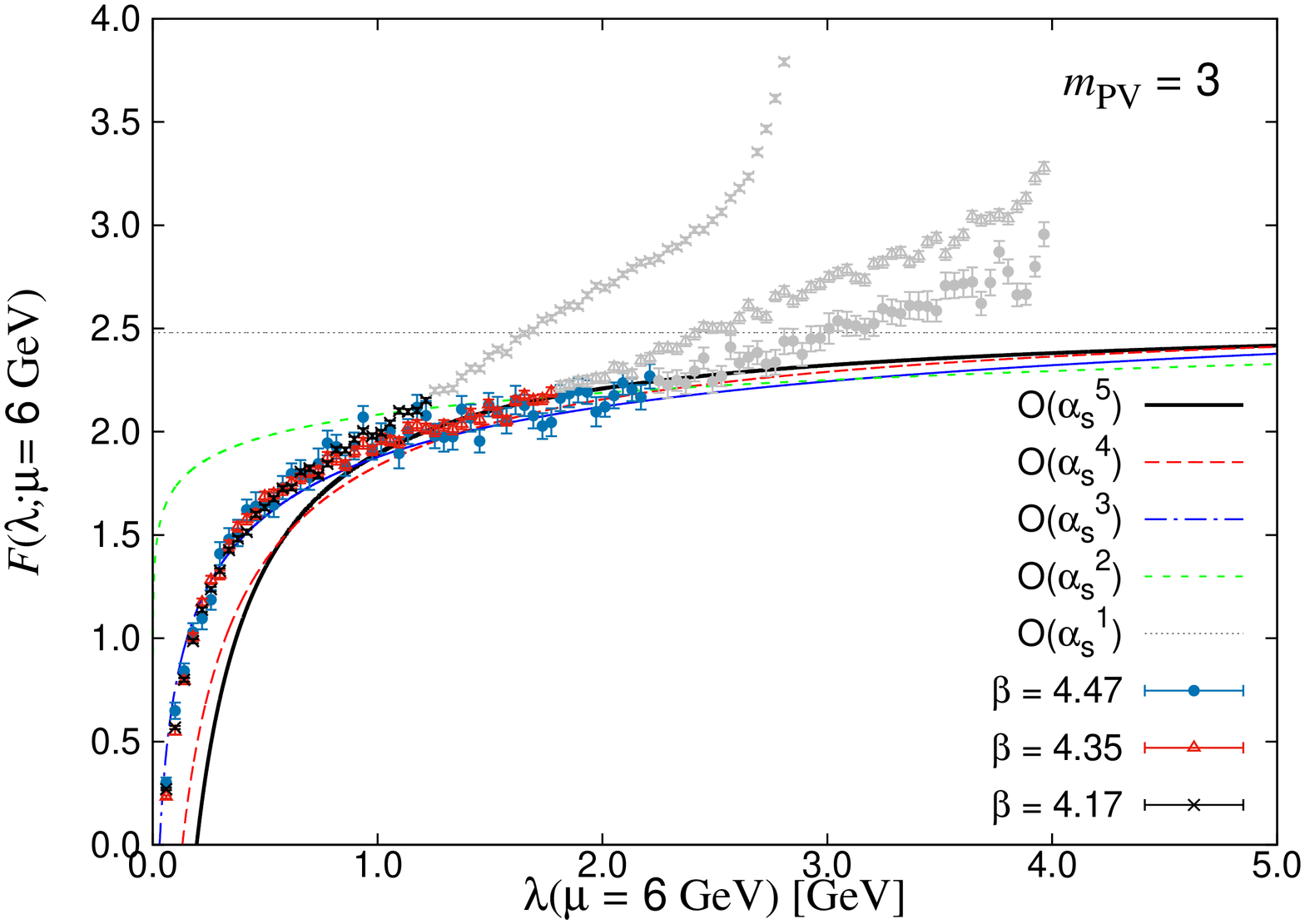}
  \includegraphics[width=6.5cm,angle=-90]{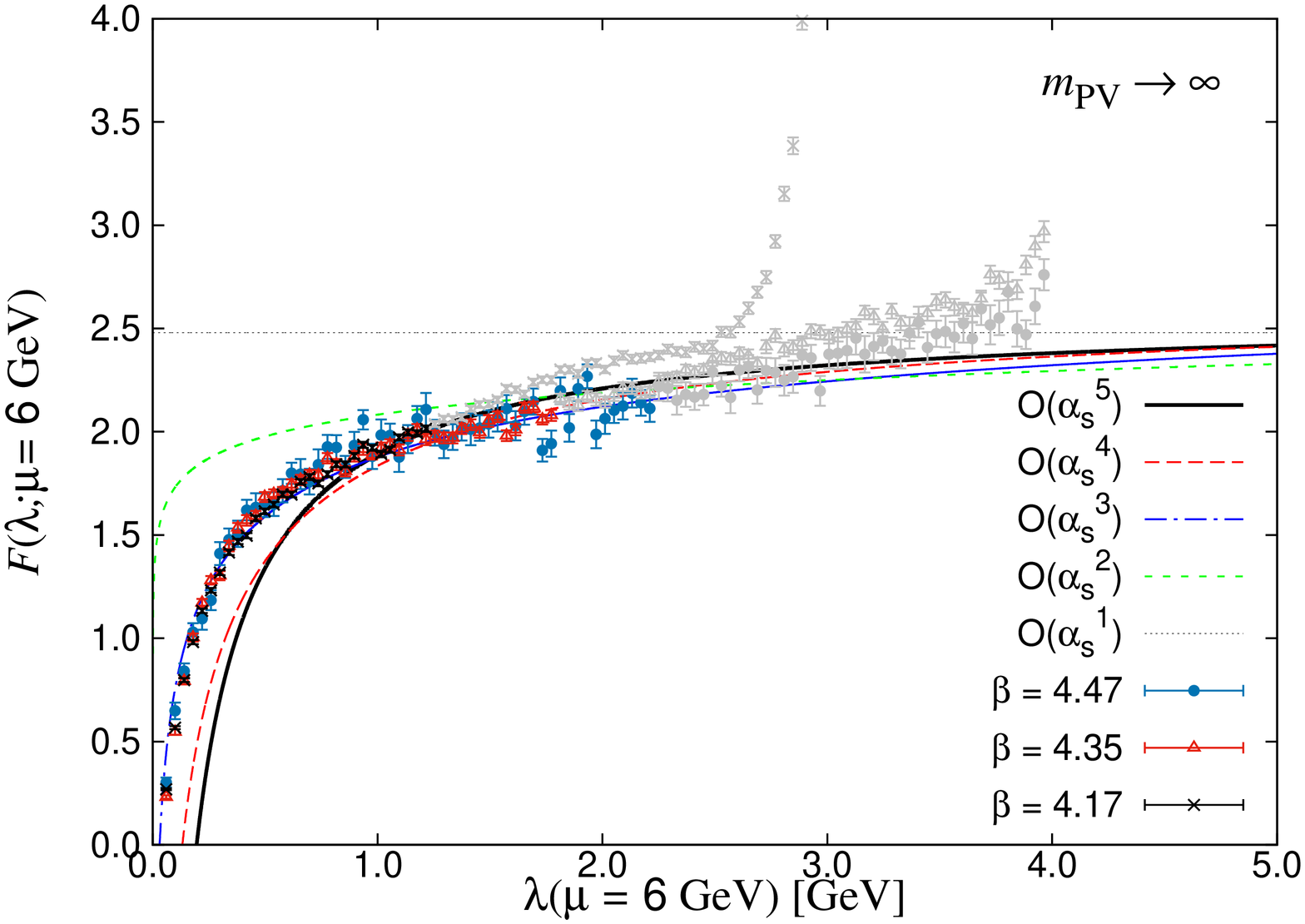}
  \caption{
    Exponent of the Dirac spectral density $F(\lambda)$ with the
    Pauli-Villars mass $am_{\mathrm{PV}}=1$ (top), $m_{\mathrm{PV}}=3$
    (middle), and $am_{\mathrm{PV}}\to\infty$ (bottom).
    The renormalization scale $\lambda(\mu)$ is set to
    $\mu$ = 6~GeV.
    The lattice data at 
    $\beta$ = 4.17 (crosses), 4.35 (triangle) and 4.47 (dots) 
    are shown.
    Grayed data points are those with $a\lambda>0.5$.
    Lines represent results from the perturbative expansion.
  }
  \label{fig:power_cont_each_beta}
\end{figure}

In Figure~\ref{fig:power_cont_each_beta} we plot $F(\lambda)$ obtained 
at each lattice spacing.
The renormalization scale is set to $\mu$ = 6~GeV.
The results at three lattice spacings are corrected 
by a factor $3/F_0(\lambda)$, derived from (\ref{eq:F_tree}), to
eliminate the leading discretization effects as discussed in
Section~\ref{sec:Tree_level}.
We find strong discretization effects even after including this correction for the leading discretization effect, especially for the standard Pauli-Villars mass, $am_{\mathrm{PV}}=1$ (top panel).
It is much improved with $am_{\mathrm{PV}}=3$ (middle panel) and with
$am_{\mathrm{PV}}\to\infty$ (bottom panel).
In the plots, the data points shown by gray symbols have $a\lambda>$
0.5, for which the discretization effects seem significant even after
the correction.

In the plots we also show results from the perturbative expansion
calculated at each order, from $O(\alpha_s)$ to $O(\alpha_s^5)$.
The strong coupling constant is taken from the world average
$\Lambda_{\mathrm{QCD}}^{(3)}$ = 332(17)~MeV for three-flavor QCD \cite{Olive:2016xmw},
and the unknown fifth-order constant is taken to be $d_5=0$ as a 
representative value.
The lattice data are transformed from $\mu$ = 2~GeV to 6~GeV using the
renormalization group equation with the same value of 
$\Lambda_{\mathrm{QCD}}^{(3)}$.
One can see that the lattice data are in good agreement with the
perturbative estimate for $am_{\mathrm{PV}}=3$ (middle panel) except for those points that are grayed out.

\begin{figure}[tbp]
  \includegraphics[width=6.5cm, angle=-90]{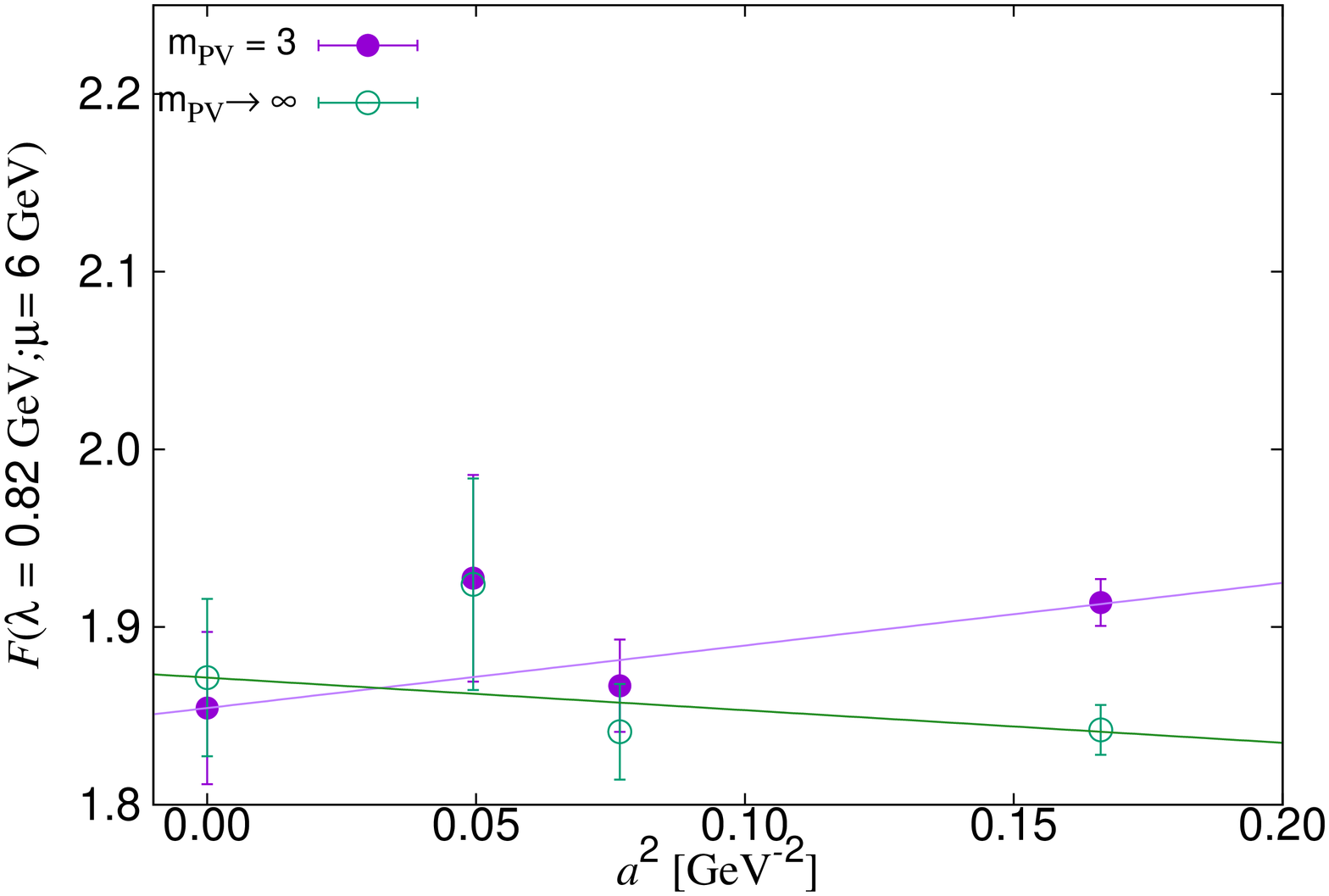}
  \includegraphics[width=6.5cm, angle=-90]{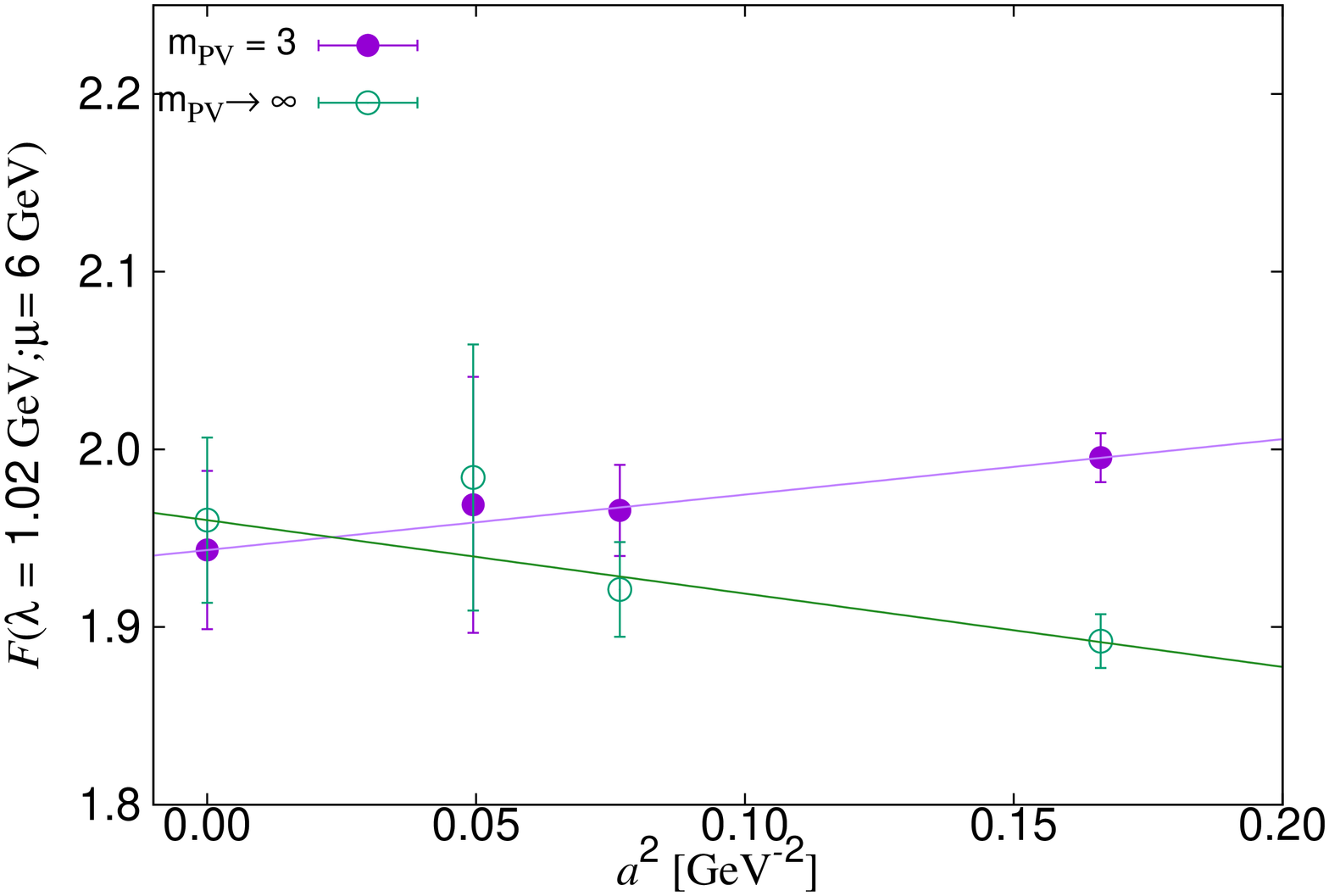}
  \includegraphics[width=6.5cm, angle=-90]{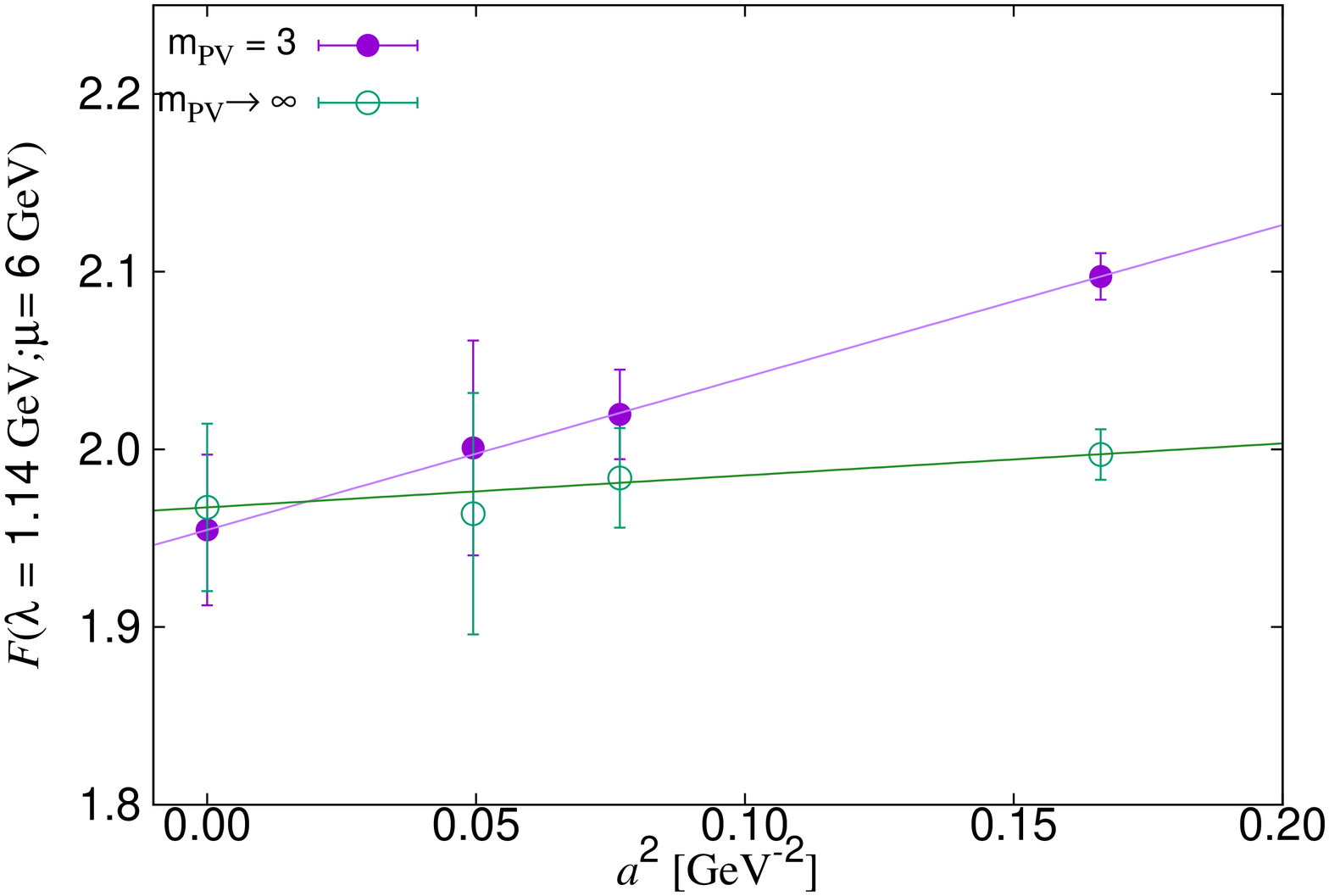}
  \caption{
    Continuum extrapolation of $F(\lambda)$ at $\lambda$ =
    0.82~GeV (top panel) 1.02~GeV (middle), 1.14~GeV (bottom).
    Results with $am_{\mathrm{PV}}=3$ (circles) and with
    $am_{\mathrm{PV}}\to\infty$ (crosses) 
    are plotted as a function of $a^2$ [GeV$^{-2}$].
  }
  \label{fig:continuum_extrap_safe}
\end{figure}

\begin{figure}[tbp]
  \includegraphics[width=6.5cm, angle=-90]{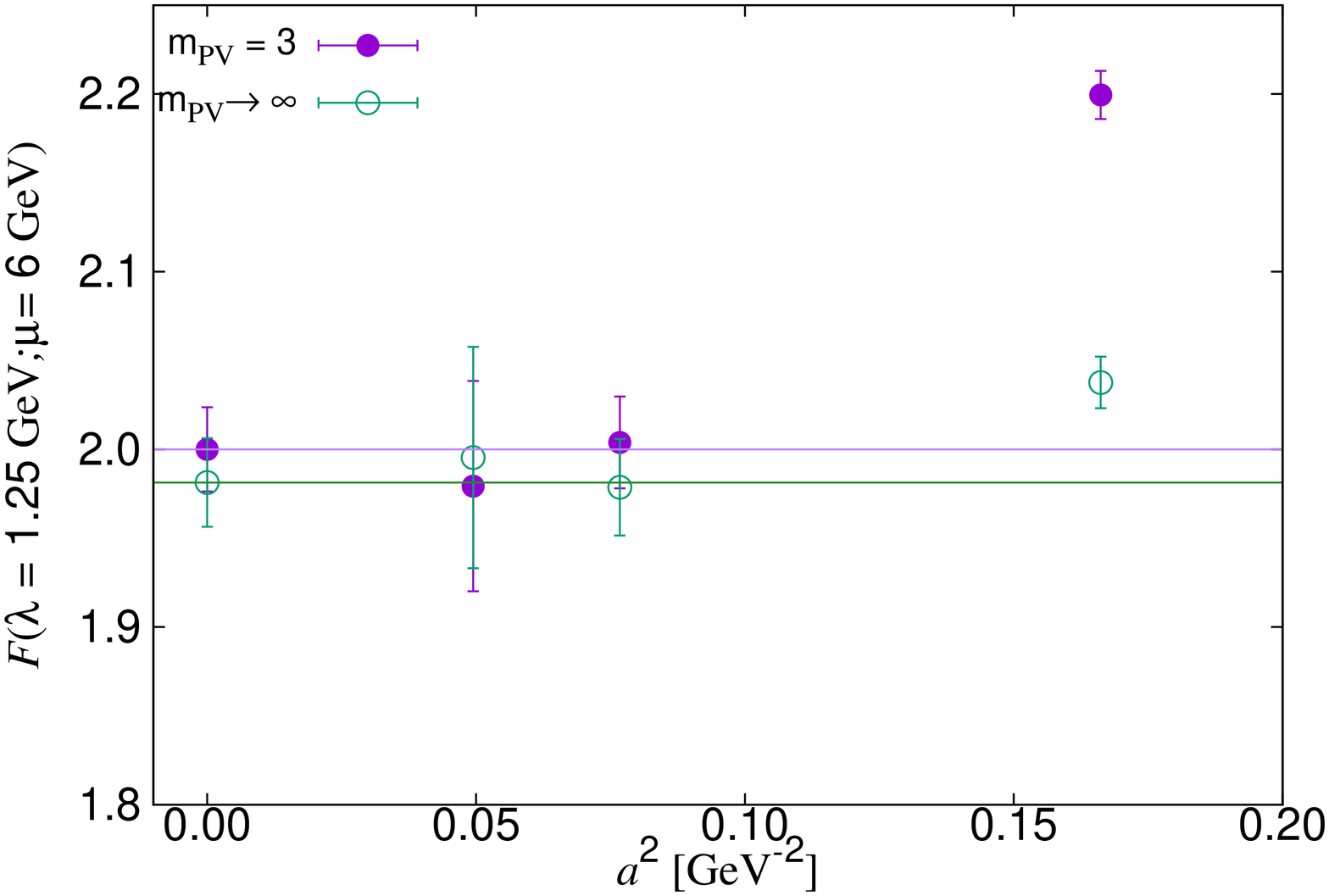}
  \caption{
    Continuum extrapolation of $F(\lambda)$ at $\lambda$ =
    1.25~GeV.
    Results with $m_{\mathrm{PV}}=3$ (circles) and with
    $m_{\mathrm{PV}}\to\infty$ (crosses)
    are plotted as a function of $a^2$ [GeV$^{-2}$].
  }
  \label{fig:continuum_extrap_bound}
\end{figure}

Before comparing the lattice results with perturbation theory in more 
detail, we extrapolate the lattice data for $F(\lambda)$ to the
continuum limit.
We choose the data at $am_{\mathrm{PV}}=3$ and
$am_{\mathrm{PV}}\to\infty$ in the following and 
extrapolate to the continuum limit assuming a linear dependence in
$a^2$. 
Figures~\ref{fig:continuum_extrap_safe} and
\ref{fig:continuum_extrap_bound}
show the continuum extrapolation for some representative values of
$\lambda$. 
With our lattice parameters, the eigenvalues at $\lambda$ = 1.22~GeV
or below satisfy the condition $a\lambda<$ 0.5 for all three lattice
spacings. 
For these bins of eigenvalues, the remaining $a^2$ dependence is well
under control as demonstrated in
Figure~\ref{fig:continuum_extrap_safe}.
The slope in $a^2$ is small and the extrapolated results of
$am_{\mathrm{PV}}=3$ and $am_{\mathrm{PV}}\to\infty$ agree well with
each other. 
This is consistent with a naive expectation of the size of 
remaining discretization effect of $O(\alpha_s a^2\lambda^2)$,
which is about 5\%.
The remaining error after the continuum extrapolation is therefore
much smaller.
For our data, it is estimated to be smaller than the statistical
error for each bin of $\lambda$.

Above $\lambda\simeq$ 1.22~GeV, the coarsest lattice fails to satisfy
the condition $a\lambda<$ 0.5.
For these bins we ignore the data from the coarsest lattice and
extrapolate the other two finer lattice data to the continuum assuming
no dependence on $a^2$, which is equivalent to taking a weighted
average. 
An example is shown in Figure~\ref{fig:continuum_extrap_bound}.
Here, the results from $am_{\mathrm{PV}}=3$ and
$am_{\mathrm{PV}}\to\infty$ agree with each other on the two finer
lattices, from which the continuum limit is estimated.
The data points at the coarsest lattice spacing, 
at around $a^2$ = 0.16~GeV$^{-2}$,
deviate substantially from the straight lines that represent the
average of the finer two lattices.

\begin{figure}[tbp]
  \includegraphics[width=9cm,angle=-90]{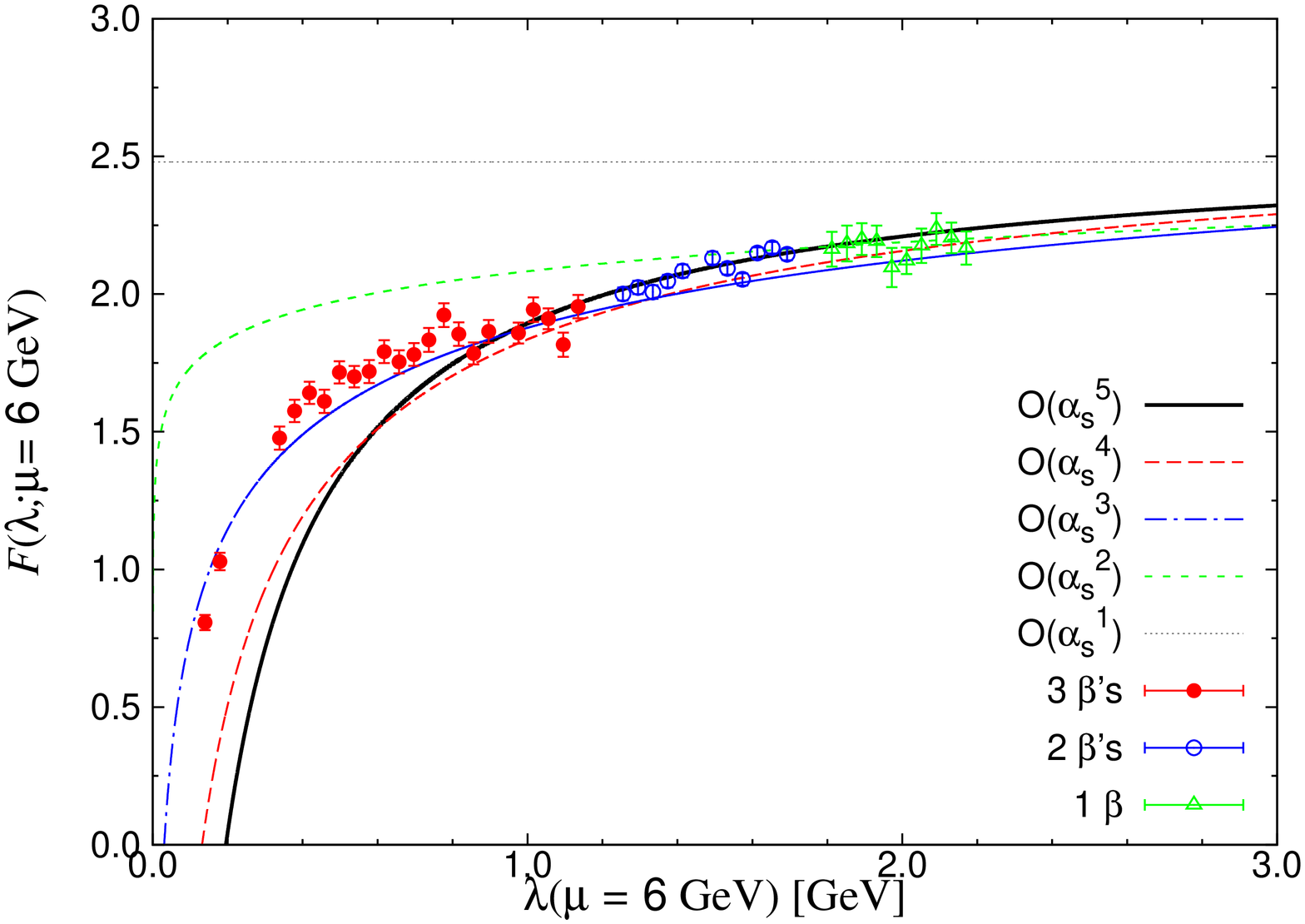}
  \caption{
    Exponent of the Dirac spectral density $F(\lambda)$ as a
    function of $\lambda(\mu=\mathrm{6~GeV})$.
    The lattice data after the continuum extrapolation are shown
    for Pauli-Villars mass $am_{\mathrm{PV}} = 3$.
    The filled circles are obtained by an extrapolation using three
    lattice spacings, while the open circles use two lattice
    spacings and green triangles are those of the finest lattice.
    Perturbative results are also drawn with truncation at various
    orders from $O(\alpha_s)$ (dotted line) to $O(\alpha_s^5)$
    (thin solid curve).
  }
\label{fig:ano_dim_lat}
\end{figure}

Figure~\ref{fig:ano_dim_lat} shows $F(\lambda)$ in the continuum
limit. 
The results from $am_{\mathrm{PV}}=3$ are plotted; 
those from $am_{\mathrm{PV}}\to\infty$ are consistent within their
statistical error.
The data points of the eigenvalue $\lambda$ below 1.22~GeV are
obtained by an extrapolation using three lattice spacings (filled
circles), and the higher eigenvalues are analyzed with only
two (open circles) or one (triangles) lattice spacings.
The highest eigenvalue we could reach in this way is 2.2~GeV.

Again in this plot we overlay the perturbative expansion of
$F(\lambda)$ up to order $\alpha_s^5$ with unknown coefficient
$d_5$ in (\ref{eq:Fmu}) set to zero.
Taking $\alpha_s(\mathrm{6~GeV})$ = 0.191, which is converted from the
world average, we find reasonable agreement between the lattice
results and perturbation theory between 
$\lambda\simeq$ 0.8~GeV and 2.2~GeV, 
which suggests that $\alpha_s(\mathrm{6~GeV})$ extracted from the
lattice data are in fair agreement with the world average.

When we compare the perturbative expansion with the lattice data, we
need to consider the non-perturbative effect that may arise as a power
correction to the spectral function.
According to the general form of the operator product expansion (OPE),
the spectral function $\rho(\lambda)$ receives a correction of the
form $\sim \langle\bar{\psi}\psi\rangle$
at the first non-trivial order.
It is suppressed by three powers of $\lambda$ relative to the leading
order contribution $\sim\lambda^3$.
This is parametrically consistent with the Banks-Casher relation
$\rho(0)=-\langle\bar{\psi}\psi\rangle/\pi$, 
which is valid in the limit of $\lambda\to 0$,
but it may not be a simple consequence of the OPE because the expansion
breaks down for small $\lambda$.
In any case, the suggested form of the power correction, {\it i.e.} a
constant in $\rho(\lambda)$, does not contribute to the exponent
$F(\lambda)$.
In other words, the power correction to $F(\lambda)$ is highly
suppressed, and the numerical result from lattice QCD,
Figure~\ref{fig:ano_dim_lat}, supports this expectation.

\section{Extraction of $\alpha_s$}
\label{sec:alpha_s}
Using the lattice results obtained for the spectral function
$\rho(\lambda)$ as an input, we attempt to extract the strong coupling 
constant $\alpha_s$. 
We use the perturbative expansion of $F(\lambda)$ up to order $\alpha_s^4$ as well as an estimate for the $O(\alpha_s^5)$ term.
The explicit formula is given in (\ref{eq:F^MS})-(\ref{eq:F^(5)}),
depending on the value of $L_\lambda=\lambda/\mu$.
We solve the equation for $\alpha_s(\mathrm{6~GeV})$
with an input $F(\lambda)$ on the right-hand side.
The error due to the unknown parameter in the $O(\alpha_s^5)$ term is
estimated as discussed below.

\begin{figure}[tbp]
  \includegraphics[width=9cm,angle=-90]{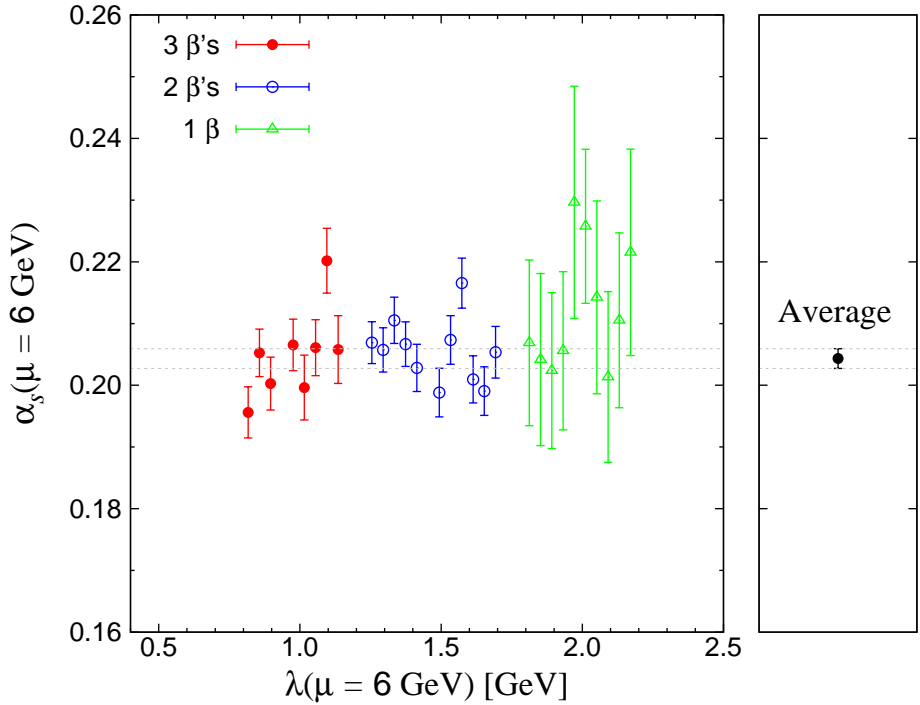}
  \caption{
    Strong coupling constant $\alpha_s(\mu=\mathrm{6~GeV})$ 
    extracted from the exponent $F(\lambda)$ of the Dirac spectral
    function $\rho(\lambda)$. 
  }
  \label{fig:alphas}
\end{figure}

The determination of $\alpha_s(\mathrm{6~GeV})$ may be done for each
value (or bin) of $\lambda$ by solving equation (\ref{eq:F^MS}).
The results are obtained as a function of $\lambda$, which is plotted
in Figure~\ref{fig:alphas}.
The results for $\alpha_s(\mathrm{6~GeV})$ are nearly independent of
$\lambda$ for $\lambda\simeq$ 0.8~GeV or higher. 
This is expected from the plot of $F(\lambda)$, 
Figure~\ref{fig:ano_dim_lat}, with which we demonstrate that the
perturbative estimate follows the lattice data down to $\lambda\simeq$
0.8~GeV. 
Below that, the perturbative expansion is expected to rapidly break
down. 
The statistical error is larger in the high energy region,
$\lambda>$ 1.8~GeV,
since the lattice results are solely from the finest lattice 
($\beta$ = 4.47) for which the statistical signal is not as good as
for other lattice spacings.
We take the central value from the bins between $\lambda$ = 0.8 and
1.25~GeV, since the data at all three lattice spacings are included in
the continuum extrapolation.
The statistically averaged value 
$\alpha_s(\mathrm{6~GeV})$ = 0.204(2) 
thus obtained is also plotted in Figure~\ref{fig:alphas} on the right
panel.  

In the following, we describe the possible sources of systematic
error and our estimates for them.

The leading discretization effects are removed by the continuum
extrapolation, and we estimate the remaining error from the difference
between the results with Pauli-Villars masses 
$am_{\mathrm{PV}}$ = 3 and $\infty$.
From an explicit calculation, we estimate the error for
$\alpha_s(\mathrm{6~GeV})$ as $\pm 0.003$.

Discretization errors may also be estimated by varying the number
of points included in the continuum extrapolation.
For the data points below $\lambda\simeq$ 1.2~GeV, for which three
$\beta$ values are included in the continuum extrapolation, we attempt
to fit the data without the coarsest point and take its difference
from the three-point extrapolation as an estimate of the systematic
error of this source.
This leads to an estimate of $\pm 0.006$ for 
$\alpha_s(\mathrm{6~GeV})$.
We therefore combine these estimates in quadrature for a conservative
estimate of discretization effects as $\pm 0.007$.

The perturbative error from unknown higher order coefficients is
estimated by allowing $\pm 50$ for $d_5$ of the $O(\alpha_s^5)$ term
(see the discussions in Section~\ref{sec:Ano_dim}).
This amounts to an uncertainty of $\pm 0.012$ for $F(\lambda)$ and 
gives a variation of $\pm 0.002$ for $\alpha_s(\mathrm{6~GeV})$. 
We also estimate the uncertainty due to the unknown $O(\alpha_s^6)$
contribution as discussed in Section~\ref{sec:Ano_dim}.
This adds another $\pm 0.052$ for $F(\lambda)$ and thus
$\pm 0.007$ for $\alpha_s(\mathrm{6~GeV})$.
Overall, the perturbative uncertainty for $\alpha_s(\mathrm{6~GeV})$
is estimated to be $\pm 0.007$.

The error due to the scale setting enters as an overall shift of the
physical scale, which affects $\alpha_s(\mathrm{6~GeV})$
due to a change of the QCD scale $\Lambda_{\mathrm{QCD}}$.
The leading scale dependence $\alpha_s(\mu)\simeq
(-2\beta_0\ln(\mu^2/\Lambda_{\mathrm{QCD}}^2))^{-1}$ implies that the
relative uncertainty 1.7\% for the input value of $t_0$ amounts to an
uncertainty of $\pm 0.001$ for $\alpha_s(\mathrm{6~GeV})$.
Similarly, the uncertainty in the renormalization scale
$Z_S(\mathrm{2~GeV})$ gives an overall shift of the eigenvalue, which
acts as a shift of the scale.
The maximal uncertainty of $Z_S(\mathrm{2~GeV})$ is 1.5\%, which leads
to $\pm 0.001$ for the estimated error in $\alpha_s(\mathrm{6~GeV})$
from this source.

\begin{table}[tbp]
  \centering
  \begin{tabular}{lr}
    \hline\hline
    statistical &     $\pm 0.002$ \\
    discretization &  $\pm 0.007$ \\
    perturbative &    $\pm 0.007$ \\
    lattice scale &   $\pm 0.001$ \\
    renormalization & $\pm 0.001$ \\
    \hline
    total &           $\pm 0.010$ \\
    \hline\hline
  \end{tabular}
  \caption{
    Estimated error for the determination of
    $\alpha_s(\mathrm{6~GeV})$ from the spectral density.
  }
  \label{tab:error_estimates}
\end{table}
 
The error estimates given above are summarized in
Table~\ref{tab:error_estimates}.
By summing them in quadrature, we estimate the total error to be 
$\pm 0.010$, which includes the statistical error.
Thus we obtain
$\alpha_s(\mathrm{6~GeV})$ = 0.204(10).
The total uncertainty is dominated by discretization effects and
higher order corrections of the perturbative expansion.

The result may be converted to the value at the $Z$ boson mass scale
$\alpha_s^{(5)}(M_Z)$ using the four-loop $\beta$ function. The threshold effect is taken into account when switching to the renormalization
scheme of four and then five dynamical quark flavors.
We obtain
$\alpha_s^{(5)}(M_Z)$ = 0.1226(36),
which may be compared to the average of the lattice results
0.1182(12) \cite{Aoki:2016frl}
(the original works contributing to this average are
\cite{Chakraborty:2014aca,McNeile:2010ji,Bazavov:2014soa,Aoki:2009tf,Maltman:2008bx})
as well as the world average of the Particle Data Group,
0.1181(11) 
\cite{Olive:2016xmw}. 
Our result is slightly higher but consistent within about one standard
deviation. 
Compared to our own result,
0.1177(26)
from the charmonium current correlator \cite{Nakayama:2016atf}, which
is obtained on the same set of lattice ensembles,
the present result is higher by about 1.1 standard deviations.


\section{Conclusions}
\label{sec:conclusion}

In this work, we perform a lattice calculation of the Dirac spectral
density in QCD using the technique to stochastically estimate the
number of eigenvalues in given intervals.
We use lattice ensembles generated with 2+1 flavors of dynamical
quarks. 
Discretization errors are subtracted as far as possible using an
estimate in the free-quark limit. The results are extrapolated to
the continuum limit from data taken at three lattice spacings.

The lattice results for the exponent of the spectral density
are in good agreement with the perturbative QCD calculation available
to order $\alpha_s^4$.
This agreement is highly non-trivial since the value of the exponent
is about 30--40\% lower than its asymptotic value and the scale (or 
the eigenvalue $\lambda$) dependence is well reproduced.
This observation adds another piece of evidence of the validity of QCD 
in both perturbative and non-perturbative regimes.

We extract the strong coupling constant using the lattice calculation 
of the spectral function as an input.
As in other determinations of $\alpha_s$, the control of
discretization effects is essential because we are working at a relatively high energy
scale.
The results are in agreement with other determinations, although the
error is not competitive.
The main reason for the relatively large uncertainty is the remaining
discretization effects.
Namely, the region of $\lambda$ is limited to 1.2~GeV or below to
fully control the continuum limit, and this energy region is not
optimal for the perturbative expansion to rapidly converge.
A lattice calculation with finer lattice spacing and/or with improved
discretization errors would be necessary to improve the precision.

\appendix
\allowdisplaybreaks[2]

\section{Perturbative formulae for Dirac spectral density}
\label{app:perturbative}%

In this appendix, we summarize useful formulae for the perturbative
calculation of the Dirac spectral density.

The $\beta$-function and the mass anomalous dimension $\gamma_m$ are
given by \cite{Baikov:2014qja,Baikov:2016tgj}
\begin{eqnarray}
  \beta(\mu) \equiv \frac{\partial\alpha_s}{\partial\ln\mu}
  & = & \beta_0\alpha_s^2
        + \beta_1\alpha_s^3
        + \beta_2\alpha_s^4
        + \beta_3\alpha_s^5
        + \beta_4\alpha_s^6
        + O(\alpha_s^7),
  \\
  \gamma_m(\mu) \equiv -\frac{\partial\ln m(\mu)}{\partial\ln\mu}
  & = & \gamma_0 \alpha_s
        + \gamma_1 \alpha_s^2
        + \gamma_2 \alpha_s^3
        + \gamma_3 \alpha_s^4
        + \gamma_4 \alpha_s^5
        + O(\alpha_s^6),
\end{eqnarray}
where the coefficients are numerically given by
\begin{eqnarray}
  \pi\beta_0 & = & -5.5 + 0.333333 n_f,
                   \nonumber\\
  \pi^2\beta_1 & = & -12.75 + 1.58333 n_f,
                   \nonumber\\
  \pi^3\beta_2 & = & -44.6406 + 8.73785 n_f - 0.188079 n_f^2,
                   \nonumber\\
  \pi^4\beta_3 & = & -228.461 + 54.2679 n_f - 3.16476 n_f^2 -
                     0.0117134 n_f^3,
                   \nonumber\\
  \pi^5\beta_4 & = & -1049.12 + 363.598 n_f - 34.312 n_f^2 
                     + 0.451714 n_f^3 + 0.00359858 n_f^4, 
\end{eqnarray}
and
\begin{eqnarray}
  \pi\gamma_0 & = & 2,
                    \nonumber\\
  \pi^2\gamma_1 & = & 8.41667 - 0.277778 n_f,
                   \nonumber\\
  \pi^3\gamma_2 & = & 39.0313 - 4.56824 n_f - 0.0540123 n_f^2,
                   \nonumber\\
  \pi^4\gamma_3 & = & 197.887 - 38.2149 n_f + 0.552325 n_f^2 +
                      0.0115864 n_f^3, 
                   \nonumber\\
  \pi^5\gamma_4 & = & 1119.41 - 287.373 n_f + 14.9648 n_f^2 
                      + 0.216637 n_f^3 - 0.000170718 n_f^4.
\end{eqnarray}

The scale dependent coefficients of the spectral function
$\rho(\lambda)$ with $\lambda\neq\mu$ is given as follows.
\begin{eqnarray}
  -\frac{\rho_1}{\pi} 
  &=& c_1 - 4 L_\lambda  \gamma_0,
      \nonumber\\
  -\frac{\rho_2}{\pi^2} 
  &=& c_2 - 2L_\lambda^2\gamma_0\left[\beta_0-4\gamma_0\right]
      + L_\lambda \left[ \beta_0 c_1 
     - 4 (c_1\gamma_0-\gamma_0^2+\gamma_1)
     \right],
     \nonumber\\
  -\frac{\rho_3}{\pi^3}
  &=& c_3 - \frac{4}{3} L_\lambda^3\gamma_0
      \left[\beta_0^2 - 6\beta_0\gamma_0 + 8\gamma_0^2\right]
      \nonumber\\
 && + L_\lambda ^2 \left[
    \beta_0^2 c_1 - 2\gamma_0(\beta_1-4c_1\gamma_0+8\gamma_0^2-8\gamma_1) + 
    \beta_0 (-6c_1\gamma_0+6\gamma_0^2-4\gamma_1)\right]
    \nonumber\\ 
 && + L_\lambda \left[
    \beta_1 c_1 + \beta_0 (2c_2-c_1\gamma_0) - 
    4 (c_2\gamma_0-c_1\gamma_0^2+\gamma_0^3+c_1\gamma_1
       - 2\gamma_0\gamma_1+\gamma_2)\right],
    \nonumber\\
  -\frac{\rho_4}{\pi^4} 
  &=& c_4 + \frac{1}{3} L_\lambda^4 \gamma_0 \left[-3 \beta_0^3 +
      22 \beta_0^2 \gamma_0 - 48 \beta_0 \gamma_0^2 + 32 \gamma_0^3\right]
      \nonumber\\
  &&+ \frac{1}{3} L_\lambda^3 \left[3 \beta_0^3c_1-2\beta_0^2
     (11 c_1 \gamma_0 - 11 \gamma_0 ^2 + 6 \gamma_1)  
     + 2 \beta_0 \gamma_0 
     (-5 \beta_1 + 24 c_1 \gamma_0 - 48 \gamma_0 ^2 + 36
     \gamma_1)\right]
     \nonumber\\
  && + \frac{1}{3} L_\lambda^3 \left[8 \gamma_0^2 (3 \beta_1 - 4 (c_1
     \gamma_0 - 3 \gamma_0^2 + 3 \gamma_1))\right]
     \nonumber\\
  && + L_\lambda^2 \left[\beta_0^2 (3 c_2 - 5 c_1 \gamma_0/2) + 
     \frac{1}{2} \beta_0 (5 \beta_1 c_1 - 
     4 (5 c_2 \gamma_0 - 7 c_1 \gamma_0^2 + 6 \gamma_0^3 + 4 c_1 \gamma_1 - 
     9 \gamma_0 \gamma_1 + 3 \gamma_2))\right]
     \nonumber\\
  && + L_\lambda^2 \left[2 (-\beta_2 \gamma_0 + \beta_1 (-3 c_1
     \gamma_0 + 3 \gamma_0^2 - 2 \gamma_1))\right]
     \nonumber\\
  && + L_\lambda^2 \left[8 (c_2 \gamma_0^2 - 2 c_1 \gamma_0^3 + 3
     \gamma_0^4 + 2 c_1 \gamma_0 \gamma_1 - 6 \gamma_0^2 \gamma_1 +
     \gamma_1^2 + 2 \gamma_0 \gamma_2))\right]
     \nonumber\\
  && + L_\lambda \left[\beta_2 c_1 + 2 \beta_1 c_2 + 3 \beta_0 c_3 -
     \beta_1 c_1 \gamma_0 - 2 \beta_0 c_2 \gamma_0 - 4 c_3 \gamma_0 +
     \beta_0 c_1 \gamma_0^2 + 4 c_2 \gamma_0^2 - 4 c_1 \gamma_0^3 + 4
     \gamma_0^4\right] 
     \nonumber\\
  && + L_\lambda \left[ - \beta_0 c_1 \gamma_1 - 4 c_2 \gamma_1 + 
     8 c_1 \gamma_0 \gamma_1 - 12 \gamma_0^2 \gamma_1 + 4 \gamma_1^2 -
     4 c_1 \gamma_2 + 8 \gamma_0 \gamma_2 - 4 \gamma_3\right],
     \nonumber\\
  -\frac{\rho_5}{\pi^5}
  &=& c_5 - \frac{4}{15} L_\lambda ^5 \gamma_0
      \left[3 \beta_0^4 - 25 \beta_0^3 \gamma_0 + 70 \beta_0^2
      \gamma_0^2 - 80 \beta_0 \gamma_0^3 + 32 \gamma_0^4 \right]
      \nonumber\\
  && + \frac{1}{3} L_\lambda^4 \left[3 \beta_0^4 c_1 + 4 \beta_0
     \gamma_0^2 (13 \beta_1 - 20 c_1 \gamma_0 + 60 \gamma_0^2 - 48
     \gamma_1) \right]
     \nonumber\\
  && + \frac{1}{3} L_\lambda^4 \left[\beta_0^3 (-25 c_1 \gamma_0 + 25
     \gamma_0^2 - 12 \gamma_1) - 16 \gamma_0^3 (3 \beta_1 - 2 c_1
     \gamma_0 + 8 \gamma_0^2 - 8 \gamma_1)\right]
     \nonumber\\
  && + \frac{1}{3} L_\lambda^4 \left[ \beta_0 ^2 \gamma_0 (-13 \beta_1
     + 70 c_1 \gamma_0 - 140 \gamma_0 ^2 + 88 \gamma_1) \right]
     \nonumber\\
  && + \frac{1}{3} L_\lambda^3 \left[\beta_0 ^3 (12 c_2 - 13 c_1
     \gamma_0) + \beta_0^2 (13 \beta_1 c_1 - 52 c_2 \gamma_0 + 88 c_1
     \gamma_0^2 - 70 \gamma_0^3 - 36 c_1 \gamma_1 + 88 \gamma_0
     \gamma_1 - 24 \gamma_2)\right]
     \nonumber\\
  && + \frac{1}{3} L_\lambda^3 \left[ - 4 \beta_0 (3 \beta_2 \gamma_0
     + \beta_1 (13 c_1 \gamma_0 - 13 \gamma_0^2 + 7 \gamma_1))
     \right]
     \nonumber\\
  && + \frac{1}{3} L_\lambda^3 \left[ -4 \beta_0 ( - 6 (3 c_2
     \gamma_0^2 - 7 c_1 \gamma_0^3 + 10 \gamma_0^4 + 5 c_1 \gamma_0
     \gamma_1 - 16 \gamma_0^2 \gamma_1 + 2 \gamma_1^2 + 4 \gamma_0
     \gamma_2)) \right]
     \nonumber\\
  && + \frac{1}{3} L_\lambda^3 \left[ -2 \gamma_0 (3 \beta_1 ^2 - 12
     \beta_1 (2 c_1 \gamma_0 - 4 \gamma_0 ^2 + 3 \gamma_1)) \right] 
     \nonumber\\
  && + \frac{1}{3} L_\lambda^3 \left[-2 \gamma_0 (4 (-3 \beta_2
     \gamma_0 + 4 (c_2 \gamma_0^2 + 3 (-c_1 \gamma_0^3 + 2 \gamma_0^4
     + c_1 \gamma_0 \gamma_1 - 4 \gamma_0^2 \gamma_1 + \gamma_1^2 + 
     \gamma_0 \gamma_2)))
     )\right]
     \nonumber\\
  && + L_\lambda^2 \left[ \frac{3 \beta_1^2 c_1}{2} + \beta_0 ^2 (6
     c_3 - 7 c_2 \gamma_0 + \frac{9 c_1 \gamma_0^2}{2} - 3 c_1 \gamma_1)
     \right]
     \nonumber\\
  && + L_\lambda^2 \left[ \beta_1 (\beta_0 (7 c_2 - 6 c_1 \gamma_0) - 
     2 (5 c_2 \gamma_0 - 7 c_1 \gamma_0^2 + 6 \gamma_0^3 + 4 c_1
     \gamma_1 - 9 \gamma_0 \gamma_1 + 3 \gamma_2))\right] 
     \nonumber\\
  && + L_\lambda ^2 \left[ \beta_0 (3 \beta_2 c_1 - 2(7 c_3 \gamma_0 -
     11 c_2 \gamma_0^2))\right] 
     \nonumber\\
  && + L_\lambda ^2 \left[- 2\beta_0  ( + 12 c_1 \gamma_0^3 - 10
     \gamma_0^4 + 6 c_2 \gamma_1 - 17 c_1 \gamma_0 \gamma_1 + 24
     \gamma_0^2 \gamma_1 - 6 \gamma_1^2 + 5 c_1 \gamma_2 - 12 \gamma_0
     \gamma_2 + 4 \gamma_3)\right]
     \nonumber\\
  && + L_\lambda ^2 \left[-2 (\beta_3 \gamma_0 + \beta_2 (3 c_1
     \gamma_0 - 3 \gamma_0^2 + 2 \gamma_1) - 4 (c_3 \gamma_0^2 - 2 c_2
     \gamma_0^3 + 3 c_1 \gamma_0^4 - 4 \gamma_0^5 
      + 2 c_2 \gamma_0 \gamma_1))\right]
     \nonumber\\
  && + L_\lambda ^2 \left[-2(-4( - 6 c_1 \gamma_0^2 \gamma_1 + 12
     \gamma_0^3 \gamma_1 + c_1 \gamma_1^2 - 6 \gamma_0 \gamma_1^2 + 2
     c_1 \gamma_0 \gamma_2 - 6 \gamma_0^2 \gamma_2 + 2 \gamma_1
     \gamma_2 + 2 \gamma_0 \gamma_3)) \right]
     \nonumber\\
  && + L_\lambda \left[ \beta_3 c_1 + 2 \beta_2 c_2 + 3 \beta_1 c_3 +
     4 \beta_0 c_4 - \beta_2 c_1 \gamma_0 - 2 \beta_1 c_2 \gamma_0 - 3
     \beta_0 c_3 \gamma_0 - 4 c_4 \gamma_0 + \beta_1 c_1 \gamma_0^2 +
     2 \beta_0 c_2 \gamma_0^2 \right]
     \nonumber\\
  && + L_\lambda \left[ 4 c_3 \gamma_0^2 - \beta_0 c_1 \gamma_0^3 - 4
     c_2 \gamma_0^3 + 4 c_1 \gamma_0^4 - 4 \gamma_0^5 - \beta_1 c_1
     \gamma_1 - 2 \beta_0 c_2 \gamma_1 - 4 c_3 \gamma_1 + 2 \beta_0
     c_1 \gamma_0 \gamma_1\right]
     \nonumber\\
  && + L_\lambda \left[ 8 c_2 \gamma_0 \gamma_1 - 12 c_1 \gamma_0^2
     \gamma_1 + 16 \gamma_0^3 \gamma_1 + 4 c_1 \gamma_1^2 - 12
     \gamma_0 \gamma_1^2 - \beta_0 c_1 \gamma_2 - 4 c_2 \gamma_2 + 8
     c_1 \gamma_0 \gamma_2 - 12 \gamma_0^2 \gamma_2 \right]
     \nonumber\\
  && + L_\lambda \left[ 8 \gamma_1 \gamma_2 - 4 c_1 \gamma_3 + 8
     \gamma_0 \gamma_3 - 4 \gamma_4 \right],
\end{eqnarray}
where $L_\lambda=\ln(\lambda/\mu)$, and 
the numerical coefficients are
\begin{eqnarray}
  c_1 & = & \frac{10}{3\pi},
            \nonumber\\
  c_2 & = & \frac{2203 - 234 n_f - 324 \pi^2 + 8 n_f \pi^2 - 48
            \zeta_3 }{72 \pi^2},
            \nonumber\\
  c_3 & = & \left\{
            \begin{array}[c]{ll}
              -5.98134 & (n_f=3)\\
              -5.90468 & (n_f=2)
            \end{array}
                         \right..
\end{eqnarray}

\section{Chirality and locality of generalized Pauli-Villars action}
\label{sec:locality_and_chirality}
The overlap-Dirac operator constructed from the domain-wall fermion
may be generalized by choosing an arbitrary value of the Pauli-Villars
mass $m_{\mathrm{PV}}$ as represented in (\ref{eq:5to4}) and
(\ref{eq:Dov}). 
In this note, we argue that this generalization does not spoil the
chirality and locality property of the overlap-Dirac operator
constructed from them.

The overlap-Dirac operator $D_{\mathrm{ov}}(m_f=0,m_{\mathrm{PV}})$ 
satisfies the relation 
\begin{equation}
  \left\{
    D_{\mathrm{ov}}^{-1}(m_f=0,m_{\mathrm{PV}}), \gamma_5
  \right\} 
  = \frac{2a \gamma_5}{(2 - (b-c)M_0)M_0m_{\mathrm{PV}}},
\end{equation}
where the numerical factor on the right-hand side depends on
$m_{\mathrm{PV}}$.
It is analogous to the similar relation for the fixed point action
$\{D_{\mathrm{FP}}^{-1},\gamma_5\}=2a\gamma_5/\kappa_f$,
where $\kappa_f$ denotes a parameter to control the block-spin
transformation \cite{Hasenfratz:1998ri}.
In the limit of $\kappa_f\to\infty$, chirality is restored while locality
is lost.
For the domain-wall fermion with a generalized Pauli-Villars mass, 
$m_{\mathrm{PV}}$ plays the same role.

The source of non-locality is in the operator
$V\equiv\gamma_5\mathrm{sgn}(\gamma_5 aD_M)=
D_M/(D_M^\dagger D_M)^{1/2}$ as well as the kernel operator $D_M$
itself. 
For the standard domain-wall fermion ($am_{\mathrm{PV}}=1$), $V$
appears only in the numerator of the definition of $aD_{\mathrm{ov}}$
given in (\ref{eq:Dov}),
and its locality is determined by that of $D_M$.
The kernel operator contains a term $1/(2+(b-c)aD_W)$, which is
exponentially localized unless the denominator develops a pole.
When $b-c=1$, the standard choice for the domain-wall fermion,
this is indeed the case as there is a mass of $O(1)$.
Once the kernel operator is known to be exponentially local, $V$ is 
also local following the argument of
\cite{Hernandez:1998et}.

\begin{figure}[tbp]
  \includegraphics[width=12cm, angle=0]{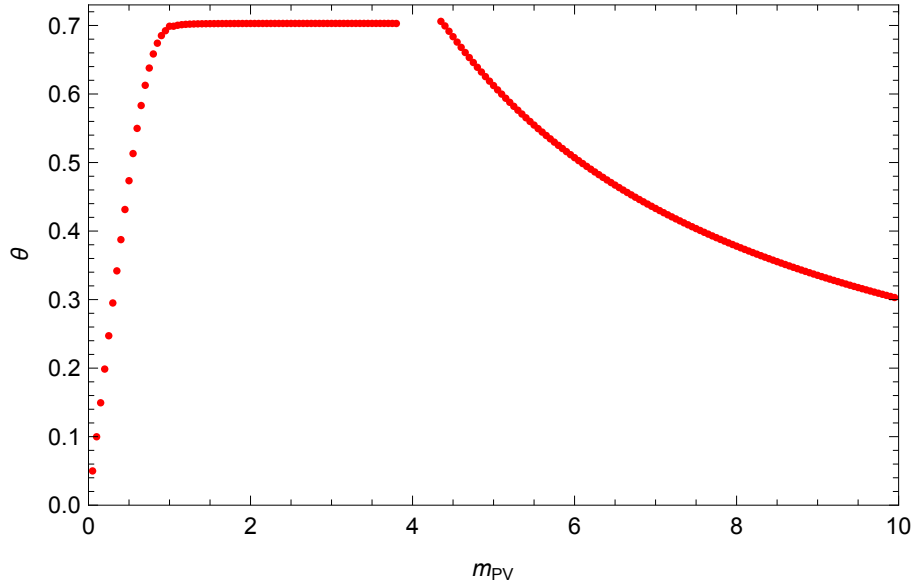}
  \caption{
    Locality of the overlap operator constructed from the domain-wall
    ferminos with different values of the Pauli-Villars mass
    $am_{\mathrm{PV}}$. 
  }
  \label{fig:pole}
\end{figure}

We can confirm this for the free field case.
We calculate the Fourier transform of the domain-wall fermion operator
and obtain its tail at long distances $|x|$.
In general, it shows an exponential fall-off $\propto e^{-\theta |x|}$.
Its rate $\theta$ is plotted as a function of $am_{\mathrm{PV}}$
in Figure~\ref{fig:pole}.

It turns out that the localization length is optimized in the range
$am_{\mathrm{PV}}\simeq$ [1,4].
Our choice $am_{\mathrm{PV}}=3$ is within this range, and it's an
equally valid choice as a domain-wall fermion formulation.

\begin{acknowledgments}
  We are grateful tothe members of the JLQCD collaboration for thir collaborative works
  to generate the data, which enable this analysis.
  We also thank B.Colquhoun for carefully reading the manuscript.
  The lattice QCD simulation has been performed on Blue Gene/Q
  supercomputer at the High Energy Accelerator Research Organization
  (KEK) under the Large Scale Simulation Program (Nos. 13/14-4,
  14/15-10, 15/16-09). 
  This work is supported in part by the Grant-in-Aid of the Japanese
  Ministry of Education (No. 25800147, 26247043, 26400259), and Grand-in-Aid for JSPS fellows.
\end{acknowledgments}



\end{document}